\documentclass[prd,showpacs,amsmath,amssymb,superscriptaddress,preprintnumbers,nofootinbib]{revtex4}
\usepackage{amssymb}
\usepackage{amsmath}
\usepackage{graphicx}

\normalbaselines
\newcommand{\nablab}{{\mathop {\rule{0pt}{0pt}{\nabla}}\limits^{\bot}}\rule{0pt}{0pt}}
\newcommand{\Fst}{{\mathop {\rule{0pt}{0pt}{F}}\limits^{\;*}}\rule{0pt}{0pt}}

\begin{document}

\title{Electrodynamics of a Cosmic Dark Fluid }

\author{Alexander B. Balakin}
\email{Alexander.Balakin@kpfu.ru}
\affiliation{Department of General Relativity and
Gravitation, Institute of Physics,Kazan Federal University, Kremlevskaya street 18, 420008, Kazan,
Russia}


\begin{abstract}
\noindent
Cosmic Dark Fluid is considered as a non-stationary medium, in which electromagnetic waves propagate, and magneto-electric field structures emerge and evolve.
A medium - type representation of the Dark Fluid allows us to involve into analysis the concepts and mathematical formalism elaborated in the framework of classical
covariant electrodynamics of continua, and to distinguish dark analogs of well-known medium-effects, such as optical activity, pyro-electricity, piezo-magnetism, electro- and magneto-striction and dynamo-optical activity.
The Dark Fluid is assumed to be formed by a duet of a Dark Matter (a pseudoscalar axionic constituent) and Dark Energy (a scalar element); respectively,
we distinguish electrodynamic effects induced by these two constituents of the Dark Fluid. The review contains discussions of ten models, which describe electrodynamic effects induced by
Dark Matter and/or Dark Energy. The models are accompanied by examples of exact solutions to the master equations, correspondingly extended; applications are considered for cosmology
and space-times with spherical and pp-wave symmetries. In these applications we focused the  attention on three main electromagnetic phenomena induced by the Dark Fluid: first, emergence of Longitudinal Magneto-Electric Clusters; second, generation of anomalous electromagnetic responses; third, formation of Dark Epochs in the Universe history.
\end{abstract}

\maketitle

\section{Introduction}

\subsection{Preface}

The term {\em Dark Fluid} was introduced into  the theory of Universe evolution in order to unify two key constitutive elements of modern cosmology: the {\em Dark Matter} and  {\em Dark Energy} (for short, DM and DE, respectively). DM and DE appeared in the scientific lexicon in
two different ways. The Dark Matter is associated foremost with the
explanation of the flat velocity curves of the spiral galaxies
rotation, and observations of a gravitational lensing (see, e.g.,  \cite{DM1,DM2,DM3,DM4} for historical details and references). The Dark Energy is considered as a reason
of the late-time accelerated expansion of the Universe discovered at the end of 20th century \cite{DE1,DE2,DE3} (see also \cite{DE4,DE5,DE6,DE7,DE8} for theories, constraints and references).
The tendency to unify DM and DE into a Dark Fluid can be motivated by two hypotheses. The first one is that such
a unified dark cosmic substratum is suitable to play the role of a dominant source of the Universe evolution, since the total contribution of the DM and DE
into the Universe energy balance is estimated to be  95\% (72 \% for DE and 23\% for DM). The second hypothesis is that DM and DE are connected by specific interactions, and links between them
display some common essence \cite{DMDE1,DMDE2,DMDE3,DMDE4,DMDE5}.
From this point of view the Dark Fluid can be considered as  an energy reservoir for the baryonic matter and cosmic photons, and thus, as a cosmic medium,
in which the electromagnetic fields of all known scales and origins are forming. There is a confidence that in a basic state the Dark Fluid does not include electrically charged particles,
however, the hypothesis exists that the Dark Fluid particles can possess electric and/or magnetic dipole moments \cite{DipoleDF}.
In other words, the Dark Fluid itself cannot be the source of electromagnetic fields, however, being in fact a cosmic medium, it can influence the processes of photon generation inside the baryonic matter, the process of photon propagation in the Universe, and the processes of organization of stationary electric and magnetic configurations. Thus, the Dark Fluid can act indirectly on photons, which bring to observers an information about the Universe evolution.

How a theorist could distinguish indirect electromagnetic effects produced by the Dark Matter from the effects produced by the Dark Energy?
For this purpose one has to choose basic models for the DM and DE. Generally, one can consider two alternative approaches (see, e.g., \cite{O1,O2,O3,O31,O4,O5,O6,O7} for review, details and references). In the framework of the first approach, the Dark Fluid is treated as a real fluid with specific equation of state; in terms of proposed extended versions of the Field Theory, in order to describe DM and DE, one has to deal with a couple of real fields: e.g., two scalar fields, the pseudoscalar and scalar fields, etc. The second (alternative) approach is based on the idea that the effect of accelerated expansion of the Universe is a cumulative result of specific interactions between gravitation and known fields: scalar, pseudoscalar, electromagnetic, massive vector, gauge, etc. The non-minimal coupling of the mentioned fields to the space-time curvature is one of the most known examples of such specific interaction (see, e.g., \cite{NM1,NM2,NM3,NM4,NM5,NM6,NM7,NM8}).

In one part of our works, we follow the idea that in terms of Field Theory the Dark Matter can be described by some pseudoscalar field, $\phi$, and  the Dark Energy can be described by (true) scalar field, $\Psi$; in terms of particle physics, this approach means that a pseudo-Goldstone boson (axion) relates to the Dark Matter, and some true boson corresponds to the Dark Energy. In the second part of our works we consider coupling of photons and pseudo-Goldstone bosons to the space-time curvature, thus replacing the real Dark Energy by some non-minimal field conglomerate, which possesses the macroscopic properties  similar to the ones typical for an effective Dark Energy.
Of course, we admit, that the pseudo-Goldstone bosons can form only one fraction of the Dark Matter (e.g., sterile neutrino can form the second fraction, etc...). In any case we consider the models, in which the DM can be coupled to the electromagnetic field only  via the pseudo-Goldstone boson subsystems.

The most probable candidates to the Dark Matter particles are axions, massive pseudo-Goldstone bosons.
The particles of this type were postulated in 1977 by Peccei and Quinn \cite{PQ} in order to solve
the problem of strong $CP$-invariance, and were introduced into
the high energy physics as new light bosons by Weinberg
\cite{Weinberg} and Wilczek \cite{Wilczek} in 1978. The description of these axions in terms of the Field Theory is
based on the introduction of a pseudoscalar field $\phi$, which is
assumed to interact with electromagnetic and $SU(2)$ - symmetric
gauge fields. The theory of interaction between electromagnetic
and pseudoscalar fields was elaborated by Ni in 1977 \cite{Ni77}. Later these
pseudo-Goldstone bosons were indicated as {\em axions}, and due to the works of Sikivie (see, e.g., \cite{Sikivie83}) we recognize
now this sector of science as {\em axion electrodynamics}.

The story, how axions appeared in the cosmological context, is also well-known.
The Dark Matter is a cosmic substance, which neither emits nor
scatters the electromagnetic radiation. The mass density distribution
of the DM is presented in the astrophysical catalogues as
a result of observations and theoretical simulations. One of the most
attractive hypothesis links the Dark Matter with massive
pseudo-Goldstone bosons, the axions, which are considered now as
an appropriate candidate for the DM  particles  (see, e.g.,
\cite{ADM1,ADM2,ADM3,ADM4,ADM5,ADM6,ADM7,ADM8,ADM9} for details, review and references). At present, the axion as a massive boson is not yet discovered; one can find the description of basic experiments, aimed for the axion detection, e.g., in \cite{AE1,AE2,AE3,AE4,AE5}.

We are interested to develop the {\em Electrodynamics} of a cosmic Dark Fluid. Clearly, the Dark Fluid produces electromagnetic effects of two types: the effects of the first one are connected with the Dark Matter, the effects of the second type are provoked by the photon coupling to the Dark Energy.
The effects of the first type are described below in the framework of axion electrodynamics and its extensions; it is the field-type representation of the axion-photon coupling.
The DF-induced electromagnetic effects of the second type, caused by the DE constituent, are described below in the framework of a macroscopic medium-type representation.

Electrodynamic systems are associated with a basic channel of information about the Universe structure, since the majority of information about Cosmos is delivered with photons registered by terrestrial and satellite detectors. There is a number of schemes for description of the DF - influence onto this channel of information. In the cosmological context, the electrodynamic system can be considered as a {\em marker}, which signalizes (by its electromagnetic response) about the variations in the state of the
Dark Fluid, the energy reservoir, into which this marker is
immersed. We are interested to answer the question: what
mechanisms might be responsible for a transmission of
information about the Dark Fluid state to the electrodynamic
systems. These mechanisms could help to reconstruct features of the Dark Fluid evolution, e.g., by
tracking down specific fine details of the spectrum of observed
electromagnetic waves, of their phase and group velocities.

Classification and discussion of schemes of mathematical description of these {\em marker-effects}
are the main purposes of the presented review. The key idea is that the Dark Fluid can be considered
as a specific {\em cosmic medium} in which the electromagnetic fields evolve. In this sense we can attract the
attention to effects, which are the analogs of well-known and well-tested effects  in the electrodynamics of moving polarizable/magnetizable
continuous media in classical (relativistic) theory. For this purpose, we introduce standardly the electromagnetic potential four-vector $A_i$, the
Maxwell tensor $F_{ik}$, which describes the electromagnetic field strength.
Next step, associated with formulation of extended equations of a Dark Fluid Electrodynamics, can be made based on formulation of an action functional.
Below we start the classification of terms included into the action functional.
The description of each model starts with the corresponding term of the total Lagrangian, then we discuss the properties of the susceptibility tensor,
and finally, consider examples of exact solutions of the master equations as illustrations.

\subsection{Ten Schemes of Description of Dark Fluid Coupling to Electromagnetic Field}

Below we consider ten models, which describe interactions between the electromagnetic field and the Dark Fluid.
Enumeration of models, displayed below in the list of models, corresponds to the Sections of the review. The models 1-4 relate to the photon couplings to the axionic Dark Matter;
the mathematical description is based on the standard axion electrodynamics (see the Model 1), and on its extensions (see the Models 2-4). The Models 5-8 describe the  coupling of photons to the DE constituent of the Dark Fluid; for the description we use the medium representation of bosonic systems associated with a scalar Dark Energy. The Model 9 deals with representation of the DE-effects as cumulative effects of non-minimal coupling of photons to the gravity field. The Model 10 presents an example of description of the DF-influence on photons mediated by coupling to a fermionic system.
More precisely, this Model 10 deals with a multi-component relativistic self-interacting plasma, and thus, it deals with the DM and DE interactions with the cooperative electromagnetic field linking charged plasma particles.

The list of models is the following.

\noindent
1. Minimal coupling of photons to the axionic Dark Matter.

\noindent
2. Non-stationary optical activity induced by the axionic Dark Matter.

\noindent
3. Gradient-type interactions with the axionic Dark Matter.

\noindent
4. Dynamo-optical interactions associated with  the axionic Dark Matter.

\noindent
5. Striction-type coupling via a scalar Dark Energy.

\noindent
6. Piezo-type coupling via a scalar Dark Energy.

\noindent
7. Pyro-type coupling via a scalar Dark Energy.

\noindent
8. Dynamo-optical interactions associated with Dark Energy.

\noindent
9. Non-minimal coupling of photons to the Dark Fluid.

\noindent
10. Electromagnetic interactions induced by the Dark Fluid in a plasma with cooperative field.

The paper is organized as follows. In Section II we consider the formalism, which gives the mathematical grounds for all ten listed models.
Sections III-XII include analysis of ten models listed above. Every Section includes discussions about the corresponding contributions into the total Lagrangian, into the master equations and into the susceptibility tensors, as well as, illustrations based on exact solutions obtained in the framework of the model under consideration.

\section{General Formalism}

\subsection{The Action Functional and Decomposition of the Total Lagrangian}

We consider the total  action functional to be of the standard form:
\begin{equation}
S_{({\rm total})} = \int d^4 x \sqrt{{{-}}g} \ \pounds \,, \quad  \pounds = \pounds_{(0)} + \pounds_{({\rm interaction})} \,.
\label{1a}
\end{equation}
Here $g$ is the determinant of the metric tensor $g_{ik}$; $\pounds$ is the total Lagrangian, which is divided into the Lagrangian of basic constituents, $\pounds_{(0)}$, and the part
$\pounds_{({\rm interaction})}$, which describes interactions of various types  between these constituents.
We include the sum of five intrinsic elements into the first part of the Lagrangian:
\begin{equation}
\pounds_{(0)} = L_{({\rm Grav})} + L_{({\rm EM})} + L_{({\rm DM})}
+ L_{({\rm DE})} + L_{({\rm OM})} \,. \label{2}
\end{equation}
The term $L_{({\rm Grav})}$ relates to the gravity field. It can be the Einstein-Hilbert Lagrangian, $L_{({\rm Grav})} \to \frac{R}{2\kappa}$ with the Ricci scalar $R$ and the Einstein
constant $\kappa {=} \frac{8 \pi G}{c^4}$; also, one can use the Lagrangian quadratic in the Riemann tensor $R^i_{\ kmn}$ and in the Ricci tensor $R_{ik}$; it can contain the Gauss-Bonnet term,... etc (see, e.g., \cite{O1,O2,O3} for review, details and references). The next term, $L_{({\rm EM})}$, describes the electromagnetic field in vacuum; since we deal with linear electrodynamics, we use the gauge-invariant term $L_{({\rm EM})} \to \frac{1}{4} F^{mn}F_{mn}$ with the Maxwell tensor $F_{mn}$. The terms $L_{({\rm DM})}$ and  $L_{({\rm DE})}$ relate to the contributions of the Dark Matter and Dark Energy, respectively; we will decode their structure later. The last term $ L_{({\rm OM})}$ describes the contribution of the so-called Ordinary Matter, in particular, the fermionic matter, which contains the contributions from electrically charged particles (electrons, positrons, protons...).

The interaction term is considered to be up to the second order with respect to the Maxwell tensor $F_{mn}$:
\begin{equation}
\pounds_{({\rm interaction})}= L_{({\rm NonEM})} + \frac12 {\cal H}^{mn} F_{mn} + \frac14 \chi^{ikmn} F_{ik}F_{mn} \,.
\label{3}
\end{equation}
The first term $L_{({\rm NonEM})}$ does not contain the Maxwell tensor, and describes all possible interactions without participation of photons; in particular, it can be the Lagrangian of interaction between DM and DE inside the Dark Fluid itself.
The part $\frac12 {\cal H}^{mn} F_{mn}$ is linear in the Maxwell tensor $F_{ik}$. The skew-symmetric tensor ${\cal H}^{mn}$ does not contain information about the electromagnetic field and describes  the so-called {\em spontaneous} polarization-magnetization. The last term $\frac14 \chi^{ikmn} F_{ik}F_{mn}$ is quadratic in $F_{mn}$; the quantity  $\chi^{ikmn}$ describes the tensor of total susceptibility.
In analogy with classical linear electrodynamics of continua (see, e.g., \cite{ED1,ED2,ED3,ED4,ED5}) we distinguish between two types of  polarization-magnetization, which can appear in the medium. The polarization-magnetization of the first type, which is indicated terminologically as spontaneous and is described by the tensor ${\cal H}^{mn}$, is produced by the forces of a non-electromagnetic origin, e.g., it can be induced by the gradient of internal temperature, stress, torsion and deformation of the medium. This term, spontaneous, appeared, first, in the theory of Phase Transition of the Second Kind, and then it has taken root in linear and non-linear electrodynamics of continua. The polarization-magnetization of the second type is produced by the influence of electromagnetic field; the corresponding contribution is described by the tensor $\chi^{ikmn} F_{mn}$, i.e., it is proportional to the Maxwell tensor. Clearly, the term susceptibility stands for coefficients, which describe the linear response of the medium to the action of the electromagnetic field.

The construction (\ref{3}) possesses the $U(1)$ symmetry; it contains the gauge invariant quantity $F_{ik}$ only. Another approach exists, which admits the appearance of gauge non-invariant terms $\nabla_k A_i{+} \nabla_i A_k$ and $\nabla_k A^k$ in the Lagrangian. For instance, in \cite{MDE} the term $\frac12 \zeta \left(\nabla_k A^k \right)^2$ is introduced into the Lagrangian, providing the extended electromagnetism to mimic the Dark Energy effects. The problem of violence of Lorentz invariance in extended equations of electromagnetism was discussed in many works; we quote for illustration only two papers \cite{LOR1,LOR2}, which give an idea how does such extensions change the results of analysis.

Our further plan is the following: first, we intend to represent the Lagrangian of interaction as a sum $\pounds_{({\rm interaction})}= L_{(1)}+ L_{(2)}+ ...+L_{(10)}$ and to specify $L_{(1)}$, $L_{(2)}$, ...$L_{(10)}$ according to the chosen models; second, for each model we plan to derive master equations for the electromagnetic field, to find and discuss the corresponding contributions to the scalar $L_{({\rm NonEM})}$ and tensors ${\cal H}^{mn}$ and $\chi^{ikmn}$; third,  for mentioned  models of interaction between DF and photons, we discuss illustrations and physical sense of  results.

\subsection{Master Equations for the Electromagnetic Field}

The Maxwell tensor $F_{ik}$ is a basic element of the $U(1)$ - symmetric covariant electrodynamics. It is standardly represented in terms of a four-vector
potential $A_i$ as
\begin{equation}
F_{ik} = \nabla_i A_{k} - \nabla_k A_{i} \,, \label{maxtensor}
\end{equation}
and thus satisfies the condition
\begin{equation}
\nabla_{k} F^{*ik} =0 \,, \label{Emaxstar}
\end{equation}
where $\nabla_k$ denotes the covariant derivative. The pseudotensor $F^{*ik} \equiv \frac{1}{2} \epsilon^{ikpq}F_{pq}$ is the
tensor dual to $F_{pq}$; the term $\epsilon^{ikpq} \equiv
\frac{1}{\sqrt{-g}} E^{ikpq}$ is the Levi-Civita (pseudo)tensor,
$E^{ikpq}$ is the absolutely skew-symmetric Levi-Civita symbol
with $E^{0123}=1$. Also, we assume that the Lorentz gauge is valid, i.e., $\nabla_k A^k=0$.

Master equations for the electromagnetic field can be derived using the variation of the action functional
(\ref{1a}) with respect to the four-vector potential $A_i$, and can be written as:
\begin{equation}
\nabla_{k} H^{ik} = - \frac{4\pi}{c} J^i \,.  \label{induc1}
\end{equation}
Here $H^{ik}$ is the induction tensor \cite{ED3}; it is convenient to represent it in the form
\begin{equation}
H^{ik} \equiv  \frac{\partial \pounds}{\partial F_{ik}} -\frac{\partial \pounds}{\partial F_{ki}} \,.
\label{4}
\end{equation}
The four-vector $J^i$ defined as
\begin{equation}
J^i = \frac{1}{4\pi} \frac{\delta L_{({\rm OM})}}{\delta A_i} \,,
\quad \nabla_i J^i = 0 \,, \label{E8}
\end{equation}
describes the electric current produced by the fermionic part of the total physical system.
We assume that $L_{({\rm OM})}$ does not include the
Maxwell tensor $F_{mn}$, nevertheless, it can depend on the
potential four-vector $A_i$, if the medium is conductive. As for the terms $L_{({\rm DM})}$ and $L_{({\rm DE})}$,
they do not contain any information about the electromagnetic field and electric charge.

\subsection{Velocity Four-Vector and Decompositions of the Maxwell and Induction Tensors}

In order to deepen the physical sense of the electrodynamic model, one needs to decompose all the quantities, introduced above,
in terms of velocity four-vector $U^i$, normalized  as
$g_{ik}U^i U^k =1$.
What is the origin of this unit four-vector? We mention only three procedures, which show, how one can introduce this four-vector.

\subsubsection{Eigen Four-Vector of the DE Stress-Energy Tensor}

The first version is associated with a search for a unit time-like eigen four-vector of the stress-energy tensor for the DE component of the Dark Fluid.
Technically, we have to introduce the DE stress-energy tensor
\begin{equation}
T^{({\rm DE})}_{ik} \equiv -\frac{2}{\sqrt{-g}} \frac{\delta[\sqrt{-g} L_{({\rm DE})}] }{\delta g^{ik}} \,,
\label{Eigen1}
\end{equation}
and then to find the eigen four-vector $U^i$ using the standard algebraic procedure
\begin{gather}
T^{({\rm DE})}_{ik} U^k = W_{({\rm DE})} U_i \,, \quad  W_{({\rm DE})} = U^i T^{({\rm DE})}_{ik} U^k \,.
\label{Eigen2}
\end{gather}
The scalar $W_{({\rm DE})}$, the corresponding eigen-value, is the DE energy-density. Based on (\ref{Eigen2}) we obtain that
\begin{gather}
T^{({\rm DE})}_{ik} = W_{({\rm DE})} U_i U_k + {\cal P}_{ik} \,,
\label{Eigen3}
\end{gather}
where the quantity ${\cal P}_{ik}$
\begin{gather}
{\cal P}_{ik} = \Delta^p_i T^{({\rm DE})}_{pq} \Delta^q_k
\label{Eigen4}
\end{gather}
is the symmetric pressure tensor, defined using the projector $\Delta^p_i \equiv \delta^p_i - U^pU_i$. Clearly, the pressure tensor satisfies the relationships
\begin{gather}
{\cal P}_{ik}U^k = \ 0 \ = {\cal P}_{ik}U^i \,,
\label{Eigen490}
\end{gather}
i.e., it is orthogonal to the velocity four-vector.

\subsubsection{Eigen Four-Vector of the DF Stress-Energy Tensor}

One can use also the same procedure for the $L_{({\rm DF})}$, which is the sum $L_{({\rm DF})}= L_{({\rm DM})}+L_{({\rm DE})}$. The corresponding formulas are similar to (\ref{Eigen2}), (\ref{Eigen3}), and we do not repeat them.

\subsubsection{Unit Dynamic Vector Field}

The third version of introduction of the unit vector field $U^i$ is associated with the so-called Einstein-aether theory (see, e.g., \cite{J1,J2,J3,J4} for details, review and references). In this theory,  a unit vector field is introduced, which is usually attributed to the velocity of an aether; in other words, this alternative theory of gravity can be indicated as vector-tensor theory of the gravitational field. Clearly, instead of aether one can consider the Dark Energy or Dark Fluid as a whole, thus introducing the four-vector of DE or DF motion.

There are other versions, but below we focus on the first and third versions only. In all these cases the velocity four-vector plays the fundamental roles in the decomposition of electrodynamic quantities. Let us start with the decomposition of the covariant derivative $\nabla_i U_k$.

\subsection{Irreducible Representations of Basic Quantities in Terms of Velocity Four-Vector  $U^i$ }

\subsubsection{Irreducible Representation of the Tensor $\nabla_i U_k$ }

The covariant derivative $\nabla_i U_k$ forms a non-symmetric tensor, which can be decomposed into a sum of
four parts, containing the acceleration four-vector $DU^{i}$,
the shear tensor $\sigma_{ik}$,
the vorticity tensor $\omega_{ik}$, and
the expansion scalar $\Theta$. This decomposition has the form
\begin{equation}
\nabla_i U_k = U_i DU_k + \sigma_{ik} + \omega_{ik} +
\frac{1}{3} \Delta_{ik} \Theta \,, \label{act3}
\end{equation}
\begin{equation}
DU_k \equiv U^m \nabla_m U_k \,, \quad \sigma_{ik}
\equiv \frac{1}{2}\Delta_i^m \Delta_k^n \left(\nabla_m U_n {+}
\nabla_n U_m \right) {-} \frac{1}{3}\Delta_{ik} \Theta  \,,
\quad  \Delta_{ik} = g_{ik}- U_i U_k \,,
\label{act49}
\end{equation}
\begin{equation}
\omega_{ik} \equiv \frac{1}{2}\Delta_i^m \Delta_k^n \left(\nabla_m
U_n {-} \nabla_n U_m \right) \,, \quad \Theta \equiv \nabla_m U^m
\,, \quad D \equiv U^i \nabla_i \,. \label{act4}
\end{equation}
The quantities $DU^{i}$, $\sigma_{ik}$, $\omega_{ik}$ and $\Delta_{ik}$ are orthogonal to the velocity four-vector $U^i$.

\subsubsection{Irreducible Representation of the Maxwell Tensor $F_{ik}$ and its Dual $F^{*}_{ik}$ }

The four-vectors of electric field $E^i$ and magnetic induction $B_i$ can be obtained from $F_{ik}$ by projections:
\begin{equation}
E^i = F^{ik} U_k \,, \quad
B_i = F^{*}_{ik} U^k \,.
\label{EB0}
\end{equation}
Clearly, $ E^i U_i =0 $ and $B_i U^i =0 $,
i.e., these four-vectors are orthogonal to the velocity four-vector $U^i$.
Based on these definitions we obtain that
\begin{equation}
F_{mn} = \delta^{pq}_{mn}E_p U_q  - \eta_{mnl} B^l \,, \quad
F^{*}_{mn} = \delta^{pq}_{mn}B_p U_q  + \eta_{mnl} E^l \,, \label{Fmn0}
\end{equation}
where $\delta^{ik}_{mn}$ is the 4-indices Kronecker
tensor
\begin{equation}
\delta^{ik}_{mn} =  \delta^{i}_{m}\delta^{k}_{n} -
\delta^{i}_{n}\delta^{k}_{m}\,,
\label{delta}
\end{equation}
and $\eta_{mnl}$ is a skew-symmetric (pseudo)tensor defined as follows:
\begin{equation}
\eta_{mnl} \equiv \epsilon_{mnls} U^s \,,
\quad
\eta^{ikl} \equiv \epsilon^{ikls} U_s \,.
\label{eta}
\end{equation}
To check the compatibility of the formulas (\ref{EB0}) and (\ref{Fmn0}) one has to keep in mind the identity
\begin{equation}
\frac{1}{2} \eta^{ikl}  \eta_{klm} = - \delta^{il}_{ms} U_l U^s
= - \Delta^i_m \,.
\label{iden2}
\end{equation}
Also, it is convenient to use the relationships
\begin{equation}
- \eta^{ikp} \eta_{mnp} = \delta^{ikl}_{mns} U_l U^s =
\Delta^i_m \Delta^k_n - \Delta^i_n \Delta^k_m \,,
\label{iden1}
\end{equation}
where
\begin{equation}
\delta^{ikl}_{mns} = \delta^{i}_{m}\delta^{k}_{n}\delta^{l}_{s}
+\delta^{k}_{m}\delta^{l}_{n}\delta^{i}_{s}
+\delta^{l}_{m}\delta^{i}_{n}\delta^{k}_{s} -
\delta^{i}_{m}\delta^{l}_{n}\delta^{k}_{s} -
\delta^{l}_{m}\delta^{k}_{n}\delta^{i}_{s} -
\delta^{k}_{m}\delta^{i}_{n}\delta^{l}_{s}
\label{i1}
\end{equation}
is the known 6-indices skew-symmetric Kronecker tensor.

\subsubsection{Irreducible Representation of the Induction Tensor}

For the Lagrangian of interaction (\ref{3}), the induction tensor (\ref{4}) is linear in the Maxwell tensor
\begin{equation}
H^{ik}  = {\cal H}^{ik}+ F^{ik} + \chi^{ikmn}F_{mn} \,.
\label{id2}
\end{equation}
Combining the second and third terms one can represent the induction tensor as
\begin{equation}
H^{ik} = {\cal H}^{ik} + {\cal C}^{ikmn} F_{mn} \,,
\label{id3}
\end{equation}
where the tensor
\begin{equation}
{\cal C}^{ikmn} \equiv \frac12 (g^{im}g^{kn}-g^{in}g^{km}) + \chi^{ikmn}
\label{id4}
\end{equation}
is known as a linear response tensor. In the framework of Lagrange approach this tensor possesses the symmetries
\begin{equation}
C^{ikmn} = C^{mnik} = - C^{kimn} = - C^{iknm} \,,\label{symmC}
\end{equation}
and thus it has 21 independent components. There is, however, another approach (see, e.g.,
\cite{ED3}), for which the symmetry $C^{ikmn} {=} C^{mnik}$ is not obligatory; this approach is faced with a new phenomenon indicated as skewon (see, e.g., \cite{S1,S2,S3} for details and references).

The four-vectors of electric induction ${\cal D}^i$ and magnetic field ${\cal H}_i$ can be obtained from $H^{ik}$ as follows:
\begin{equation}
{\cal D}^i = H^{ik} U_k \,, \quad
{\cal H}_i = H^{*}_{ik} U^k \,.
\label{EB}
\end{equation}
Again, one can see that ${\cal D}^i U_i =0$ and ${\cal H}_i U^i =0 $. For the induction tensor there exist decompositions
\begin{equation}
H_{mn} = \delta^{pq}_{mn}{\cal D}_p U_q  - \eta_{mnl} {\cal H}^l \,, \quad
H^{*}_{mn} = \delta^{pq}_{mn}{\cal H}_p U_q  + \eta_{mnl} {\cal D}^l \,, \label{Fmn}
\end{equation}
which are similar to the ones for the Maxwell tensor (\ref{Fmn0}).

\subsubsection{Irreducible Representation of the Tensor of Spontaneous Polarization-Magnetization ${\cal H}^{ik}$}

The skew-symmetric tensor ${\cal H}^{ik}$
can be decomposed as (see, e.g., \cite{ED1})
\begin{equation}
{\cal H}^{ik} =  \delta^{ik}_{mn}U^n {\cal P}^{m} -
\epsilon^{ikmn}U_n {\cal M}_{m}  \,,
\label{P0}
\end{equation}
where ${\cal P}^{m}$ is the polarization four-vector and
${\cal M}^{m}$ is the magnetization
pseudo four-vector. Inverting the relation (\ref{P0}) we find
\begin{equation}
{\cal P}^{i} \equiv {\cal H}^{ik}U_k  \,,
\quad
{\cal M}^{i} \equiv \frac{1}{2} \epsilon^{ikmn}{\cal H}_{mn}U_k \,,
 \label{M111}
\end{equation}
with
${\cal P}^{i} U_i  = 0$, ${\cal M}_{i} U^i = 0$.

\subsubsection{Irreducible Representation of the Linear Response Tensor  $C^{ikmn}$}

In terms of the velocity four-vector $U^i$ the tensor $C^{ikmn}$  can be reconstructed as follows:
\begin{eqnarray}
C^{ikmn} &=& \frac12 \left[
\varepsilon^{im} U^k U^n - \varepsilon^{in} U^k U^m +
\varepsilon^{kn} U^i U^m - \varepsilon^{km} U^i U^n \right]- \nonumber\\
&&-\frac12
\eta^{ikl}(\mu^{-1})_{ls}  \eta^{mns}  - \nonumber\\
&& -\frac12 \left[\eta^{ikl}(U^m\nu_{l \ }^{\ n} - U^n \nu_{l
\ }^{\ m}) + \eta^{lmn}(U^i \nu_{l \ }^{\ k} - U^k
\nu_{l \ }^{\ i} ) \right] . \label{Cdecomposition}
\end{eqnarray}
Here $\varepsilon^{im}$, $(\mu^{-1})_{pq}$ and $\nu_{p \ }^{\ m}$
are defined as
\begin{eqnarray}
\varepsilon^{im} &=& 2 C^{ikmn} U_k U_n\ , \nonumber\\
(\mu^{-1})_{pq} &=& - \frac{1}{2} \eta_{pik} C^{ikmn} \eta_{mnq}\ , \nonumber\\
\nu_{p \ }^{\ m} &=& \eta_{pik} C^{ikmn} U_n \ .
\label{emunu}
\end{eqnarray}
The tensors $\varepsilon_{ik}$ and $(\mu^{-1})_{ik}$ are
symmetric, but $\nu_{l \ }^{\ k}$ is non-symmetric. These tensors are
orthogonal to the velocity four-vector $U^i$,
\begin{equation}
\varepsilon_{ik} U^k = 0\ , \quad (\mu^{-1})_{ik} U^k = 0\ , \quad
\nu_{l \ }^{\ k} U^l = 0 = \nu_{l \ }^{\ k} U_k\ .
\label{ortho2}
\end{equation}
The definitions (\ref{EB}) and
the decomposition (\ref{Cdecomposition}) give
\begin{equation}
{\cal D}^i = \varepsilon^{im} E_m - B^l \nu_{l}^{\ i}\  \quad
{\rm and} \quad {\cal H}_i = (\mu^{-1})_{im} B^m  + \nu_{i}^{\ m}
E_m  \,. \label{linlawDH}
\end{equation}
The quantity
$\varepsilon^{im}$ is a four-dimensional analog of the dielectric permittivity tensor; $\mu_{pq}$ is a four-dimensional analog of the
magnetic permeability tensor; the quantity $\nu_{p}^{\ m}$ describes
magneto-electric cross effects \cite{ED1,ED3,ED4}.
The 21 independent components of
$C^{ikmn}$ include 6 components of $\varepsilon^{im}$,
6 components of $(\mu^{-1})_{pq}$ and 9 components of
$\nu_{p}^{\ m}$.
The trace of the tensor $C^{ikmn}$ can be reduced to the traces of permittivity tensors
\begin{equation}
C^{ikmn} g_{im}g_{kn} = \varepsilon^k_k +  ( \mu^{-1})^k_{k}  \,.
\label{spur}
\end{equation}
The trace of the magneto-electric tensor can be expressed as
\begin{equation}
\nu_{k \ }^{\ k} = - \frac14 \epsilon_{ikmn} C^{ikmn}  \,.
\label{spur2}
\end{equation}
The so-called Post constraint $\nu_{k \ }^{\ k}=0$ (see, e.g., \cite{Post} for details) provides the tensor $C^{ikmn}$ to possess 20 independent components instead of 21.

\subsubsection{Reduction of Electrodynamic Equations}

In terms of quantities $E^i$, $B^i$, $ {\cal D}^i$ and ${\cal D}^i$ the
current-free Maxwell equations can be rewritten in the form, which is well-known in classical electrodynamics.

\noindent
First, the Gauss law takes the form
\begin{equation}
\nablab_k {\cal D}^k =  \omega_k {\cal H}^k \,.
\label{EHDB51}
\end{equation}
Second, the law of the magnetic flux conservation is of the form
\begin{equation}
\nablab_k B^k = - \omega_k E^k \,.
\label{EHDB4}
\end{equation}
Third, the Amp\`ere law can be written as
\begin{equation}
\Delta^{ik} D {\cal D}_k - \eta^{ikm} \nablab_k {\cal H}_m = - 2
\Delta^i_k {\cal H}_m \omega^{*km} + \left(\sigma^{ik} {-}
\omega^{ik} {-} \frac23 \Theta \Delta^{ik} \right) {\cal D}_k \,.
\label{EHDB52}
\end{equation}
Fourth, the Faraday law is
\begin{equation}
\Delta^{ik} DB_k + \eta^{ikm} \nablab_k E_m = 2 \Delta^i_k E_m
\omega^{*km} + \left(\sigma^{ik} {-} \omega^{ik} {-} \frac23
\Theta \Delta^{ik} \right)B_k \,. \label{EHDB5}
\end{equation}
Here we used the standard definition
$\omega^i \equiv {-} \eta^{ikm} \nabla_k U_m$ for the local
angular  velocity of rotation, and the definition $\nablab_k = \Delta_k^i \nabla_i$ for the operator of spatial gradient.

\subsection{Gravity Field Equations}

\subsubsection{The Structure of Master Equations for the Gravity Field}

In order to obtain the equations of the gravitational field one has to find the variation of the action functional (\ref{1a}) with respect to
metric. For the case, when one uses the Einstein-Hilbert Lagrangian for the gravity field, the master equations can
be written in the following form
\begin{equation}
R_{ik}-\frac{1}{2}Rg_{ik} = \kappa \left[ T^{(0)}_{ik} + T^{({\rm
interaction})}_{ik}\right] \,. \label{EineqMIN}
\end{equation}
The first quantity,  $T^{(0)}_{ik}$, formally obtained as
\begin{equation}
T^{(0)}_{ik} = - \frac{2}{\sqrt{-g}} \frac{\delta
\left[\sqrt{-g} \ \pounds_{(0)}\right]}{\delta g^{ik}} \,,
\label{TAX1}
\end{equation}
contains the sum of four tensors
\begin{equation}
T^{(0)}_{ik} = T^{({\rm EM})}_{ik} + T^{({\rm DM})}_{ik} + T^{({\rm DE})}_{ik} + T^{({\rm OM})}_{ik} \,,
\label{TAX2}
\end{equation}
with evident indication. The first addend is the standard stress-energy tensor of the vacuum electromagnetic field
\begin{equation}
T^{({\rm EM})}_{ik} = \frac14 g_{ik} F_{mn} F^{mn}-F_{im}F_k^{\ m} \,.
\label{TAX3}
\end{equation}
The stress-energy tensor of the Dark Energy, $T^{({\rm DE})}_{ik}$, is presented by (\ref{Eigen3}). The stress-energy tensor of the axionic Dark Matter, $T^{({\rm DM})}_{ik}$, will be discussed in the next Subsection.
The algebraic decomposition of the stress-energy tensor of ordinary matter, $T^{({\rm OM})}_{ik}$, is more sophisticated than (\ref{Eigen3}), since the macroscopic
velocity four-vector $U^i$ is already fixed as an eigen four-vector of
the DE stress-energy tensor. The term $T^{({\rm OM})}_{ik}$ is of the form
\begin{equation}
T^{({\rm OM})}_{ik} \equiv W^{({\rm OM})} U_i U_k + U_i I_k^{({\rm
OM})} + U_k I_i^{({\rm OM})} + P^{({\rm OM})}_{ik}\,, \label{Tmatter}
\end{equation}
and includes the heat-flux four-vector $I_k^{({\rm OM})} \equiv
\Delta^l_k T^{({\rm OM})}_{ls} U^s$.

The stress-energy tensor
\begin{equation}
T^{({\rm interaction})}_{ik} = - \frac{2}{\sqrt{-g}} \frac{\delta
\left[\sqrt{-g} \ \pounds_{({\rm interaction})}\right]}{\delta g^{ik}}
\label{TAX22}
\end{equation}
is originated from the interaction terms; its structure will be considered in the next Sections.

\subsubsection{Stress-Energy Tensor of the Dark Matter}

The Lagrangian of the axionic Dark Matter  presented in terms of dimensionless pseudoscalar field $\phi$, is chosen to be of the form
\begin{equation}
L_{({\rm DM})} = - \frac{1}{2} \Psi^2_0 \left[\xi  \nabla^m \phi \nabla_m \phi -
V(\phi^2)\right]  \,.
\label{TAX4}
\end{equation}
The corresponding stress-energy tensor reads
\begin{equation}
T_{ik}^{({\rm DM})} = \Psi^2_0 \left\{\xi \nabla_i \phi \nabla_k
\phi -  \frac{1}{2} g_{ik} \left[ \xi \nabla^m \phi \nabla_m \phi -
V(\phi^2)\right] \right\} \,.
\label{TAX5}
\end{equation}
Here the constant $\Psi_0 = 1/g_{A\gamma \gamma}$ describes the inverted constant of the axion-photon coupling, $g_{A\gamma \gamma}$ (see, e.g., \cite{ADM1,ADM5,ADM7}).
The parameter $\xi$ takes two values: $\xi{=}1$ and $\xi{=}-1$. The first case corresponds to the standard (canonic) pseudoscalar field; when $\xi{=}-1$,
one deals with phantom-like pseudoscalar field, or in other words, the pseudoscalar field with negative kinetic term (see, e.g., \cite{Sushkov} for discussion of similar idea in terms of scalar fields). As usual, $V(\phi^2)$ is the potential of this pseudoscalar field.

There is an alternative description of the stress-energy tensor of the axionic Dark Matter, it has the form typical for the fluid-type representation:
\begin{equation}
T_{ik}^{({\rm DM})}= W^{({\rm DM})} U_i U_k + I_i U_k + I_k U_i + P^{({\rm DM})}_{ik} \,.
\label{TAX55}
\end{equation}
Comparing (\ref{TAX5}) with (\ref{TAX55}), one can find the following.
First, an axionic system, considered as a fluid, is characterized by  a heat-flux four-vector
\begin{equation}
I^i \equiv  U^p \ T^{({\rm DM})}_{pq} \ \Delta^{iq} = \xi \Psi^2_0 \ D\phi \ \nablab^i \phi \,.
\label{q1}
\end{equation}
The energy-density scalar $W^{({\rm DM})}$ can be written as follows
\begin{equation}
W^{({\rm DM})} \equiv U^p T_{pq}^{({\rm DM})}U^q = \frac{1}{2} \Psi^2_0 \left[\xi (D\phi)^2 + V - \xi \nablab^i \phi \nablab_i \phi \right]\,.
\label{cl13}
\end{equation}
The tensor of the DM pressure  has the form
$$
 P^{({\rm DM})}_{ik} \equiv \Delta^p_i T_{pq}^{({\rm DM})}\Delta^q_k =
 $$
 \begin{equation}
 = \frac{1}{2} \Psi^2_0 \Delta_{ik} \left[V{-}\xi(D\phi)^2 {-} \xi \nablab^m \phi \nablab_m \phi \right] + \xi \Psi^2_0 \nablab_i \phi \nablab_k \phi \,.
\label{cl14}
\end{equation}
For the canonic pseudoscalar field ($\xi=1$),  we deal with bilingual description of the axionic Dark Matter with the following compliance:
\begin{equation}
\Psi^2_0 (D \phi)^2 = \frac12 \left[W^{({\rm DM})} + P^{({\rm DM})} \right]\left[1+ \sqrt{1-{\cal Q}^2} \right] \,,
\label{bili1}
\end{equation}
\begin{equation}
\Psi^2_0 V(\phi^2) = \left[W^{({\rm DM})} + P^{({\rm DM})} \right]\left[1+ \sqrt{1-{\cal Q}^2} \right] - 2 P^{({\rm DM})} \,,
\label{bili2}
\end{equation}
\begin{equation}
\Psi_0 \nablab^i \phi = I^i \left\{\frac12 \left[W^{({\rm DM})} + P^{({\rm DM})} \right]\left[1+ \sqrt{1-{\cal Q}^2} \right]\right\}^{-\frac12} \,.
\label{bili3}
\end{equation}
Here we used the following definitions:
\begin{equation}
P^{({\rm DM})} \equiv - \frac13 \Delta^{ik} P^{({\rm DM})}_{ik}  \,, \quad {\cal Q} \equiv \frac{2}{\sqrt3} \frac{I}{\left[W^{({\rm DM})} + P^{({\rm DM})}\right]} \,, \quad
I^2 = - I^i I_i >0 \,.
\label{bili4}
\end{equation}
When the axionic Dark Matter is considered as a homogeneous substance, i.e., $\nablab_i \phi=0$, we obtain that the effective heat-flux four vector vanishes, $I^i=0$, and thus ${\cal Q}=0$.
Then the formulas (\ref{bili1}) and (\ref{bili2}) cover the known result
\begin{equation}
\Psi^2_0 (D \phi)^2 =  \left[W^{({\rm DM})} + P^{({\rm DM})} \right] \,, \quad \Psi^2_0 V(\phi^2) = \left[W^{({\rm DM})} - P^{({\rm DM})} \right]  \,,
\label{bili11}
\end{equation}
which was used, e.g., in \cite{O1} in the context of study of the scalar field evolution.

\subsection{Master Equation for the Axion Field}

The evolutionary equation for the pseudoscalar field is the result of variation of the total action functional with respect to $\phi$. This equation
\begin{equation}
\left[\xi \nabla^m \nabla_m  + V^{\prime}(\phi^2)\right]\phi  = {\cal J} \,, \quad  {\cal J} \equiv   - \frac{1}{\Psi^2_0}{\frac{\delta  }{\delta \phi}} \pounds_{({\rm interaction})}
\label{phi1}
\end{equation}
depends essentially on the structure of the Lagrangian of interactions, and we return to this equation below by fixing the model assumptions.

\section{Model 1. Minimal Coupling of Photons to the Axionic Dark Matter}

In \cite{Ni77}  Wei-Tou Ni has introduced a new interaction term into the Lagrangian of the electromagnetic theory; the corresponding part of the Lagrangian was of the form:
\begin{equation}
L_{(1)} = \frac14 \phi F^{*}_{mn}F^{mn} \,.
\label{1model1}
\end{equation}
Since $\phi$ is a pseudoscalar field, the product $\phi F^{*}_{mn}$ is the true tensor.
We add the term $L_{(1)}$ into $\pounds_{({\rm interaction})}$, and consider the corresponding contributions only.

\subsection{Basic Quantities and Equations}

The coupling term (\ref{1model1}) introduces the contribution $H^{ik}_{(1)} = \phi F^{*ik}$ into the total induction tensor, and, consequently, the term
\begin{equation}
\chi^{ikmn}_{(1)} = \frac12 \phi \epsilon^{ikmn}
\label{1model2}
\end{equation}
into the total susceptibility tensor. Clearly, this term does not contribute to the dielectric and magnetic permittivity tensors $\varepsilon_{ik}$ and $(\mu^{-1})_{mn}$,
and only the contribution into the tensor of magneto-electric cross effects is nontrivial:
\begin{equation}
\nu^{lm}_{(1)}= -\phi \Delta^{lm}\,.
\label{1model3}
\end{equation}
The corresponding contribution into the stress-energy tensor gives trivial result, since
\begin{equation}
T^{(1)}_{pq} = - \frac{2}{\sqrt{-g}} \frac{\delta
\left[\sqrt{-g} \phi \epsilon^{ikmn}F_{ik}F_{mn} \right]}{\delta g^{pq}} = - \frac{8}{\sqrt{-g}} \frac{\delta
\left[\phi E^{ikmn}\partial_i A_k \partial_m A_n \right]}{\delta g^{pq}} =0 \,.
\label{1model4}
\end{equation}
The pseudoscalar source in the right-hand side of the equation for the axion field (\ref{phi1}) is of the form
\begin{equation}
{\cal J} \ \to \  - \frac{1}{4 \Psi^2_0} F^{*}_{mn} F^{mn} \,.
\label{1model5}
\end{equation}
With the minimal coupling term $L_{(1)}$, the electrodynamic equations for a non-conducting medium have the form
\begin{equation}
\nabla_k [F^{ik} + \phi F^{*ik}] = 0 \ \rightarrow \  \nabla_k F^{ik} = -  F^{*ik} \nabla_k \phi \,.
\label{1model6}
\end{equation}
The simplification of these equations is the consequence of the relationship $\nabla_k F^{*ik}=0$; the last representation of the master equations of the axion electrodynamics shows explicitly, that axionic Dark Matter influences the photons if and only if the gradient four-vector $\nabla_i \phi$ is not equal to zero, i.e.,
when the pseudoscalar (axion) field is inhomogeneous and/or non-stationary. Clearly, this gradient four-vector can be
time-like ($\nabla_i \phi \nabla^i \phi>0$), space-like ($\nabla_i \phi \nabla^i \phi<0$), and null ($\nabla_i \phi \nabla^i \phi=0$). These three cases can be illustrated by models with relic cosmological axions, axions distributed around spherically symmetric static objects, and  axions in a gravitational pp-wave field, respectively.
Below we present new results, which we obtained for two models from the mentioned three ones.

\subsection{Relic Cosmological Axions, Cold Dark Matter and Terrestrial Magnetic and Electric Fields}

Let us assume that the gravitational background is given, the
space-time is of the Friedmann - Lema\^itre - Robertson -
Walker (FLRW) type
\begin{equation}
ds^2 = a^2(x^0)\left[ (dx^0)^2 - dl^2 \right] \,, \label{metric1}
\end{equation}
with the scale factor $a$ and Hubble function $H = \frac{1}{a^2} \frac{d a}{d x^0}$.
Time parameter $t$ is connected with $x^0$ by the differential relation
$a(x^0) \ dx^0 = dt $; below the dot relates to the derivative with respect to time $t$.
The metric in the three-space is represented in the spherical coordinates
\begin{equation}
dl^2 = dr^2 + r^2(d\theta^2 + \sin^2{\theta} d \varphi^2)\,,
\label{metric12}
\end{equation}
and  $\sqrt{-g}= a^4 \ r^2 \sin{\theta}$. We do not consider backreaction of the electromagnetic
field on the gravitational field and neglect the direct
influence of the terrestrial gravity field on the electric and
geomagnetic fields in comparison with the influence of the relic
cosmological axions. In other words, we assume that the number of axions produced by the
macroscopic electromagnetic field $F_{ik}$ is much less than the
number of relic (primordial) axions created in the early Universe.
This means that we neglect the electromagnetic source in the
right-hand side of the equation (\ref{phi1}) and consider the
function $\phi(t)$ to satisfy the decoupled equation
\begin{equation}
\ddot{\phi} + 3H \dot{\phi} +  \mu^2 \phi = 0 \,.
\label{ax2}
\end{equation}
For the cold Dark Matter with $P^{({\rm DM})}\to 0$ and $W^{({\rm DM})} \to \rho_{({\rm DM})}$ the relationships (\ref{bili11}) yield
\begin{equation}
q \equiv a \ \dot{\phi} \to  \pm \frac{a}{\Psi_0} \sqrt{\rho_{({\rm DM})}} \,,
\label{cl154}
\end{equation}
where $\rho_{({\rm DM})}$ is the mass density of the cold Dark Matter. In \cite{BG2013} we have solved the electrodynamic equations
(\ref{1model6}) in the background space-time with metric (\ref{metric1}), (\ref{metric12}), for the axion field with $q=const$ (see (\ref{cl154}));
below we briefly discuss the main results of this work.

\subsubsection{Axion Magnetostatics}

Searching for radial, meridional and azimuthal components of the static terrestrial magnetic field
\begin{equation}
B_{({\rm rad})} = - \frac{1}{r^2 \sin{\theta}} \frac{\partial A_{\varphi}}{\partial \theta}   \,,
\quad B_{({\rm merid})} =  - \frac{1}{r \sin{\theta}} \frac{\partial A_{\varphi}}{\partial r} \,,
\quad
B_{({\rm azim})}  = -\frac{1}{r} \left[ \frac{\partial A_{\theta}}{\partial r} - \frac{\partial A_{r}}{\partial \theta}  \right] \,,
\label{ms5}
\end{equation}
as solutions to the equations of axion electrodynamics (\ref{1model6}),
we have found in \cite{BG2013}  that
\begin{equation}
B_{({\rm rad})} = - \sum_{n=1}^{\infty} \frac{n(n{+}1)}{r}  P_n(\cos{\theta}) \ \Re_{n}(r,q) \,,
\label{ms15}
\end{equation}
\begin{equation}
B_{({\rm merid})} = - \sum_{n=1}^{\infty} P^{(1)}_{n}(\cos{\theta}) \frac{1}{r} \frac{d}{dr} \left[r \Re_n(r,q) \right]
 \,,
\label{ms16}
\end{equation}
\begin{equation}
B_{({\rm azim})} =  - q \sum_{n=1}^{\infty} P^{(1)}_{n}(\cos{\theta}) \Re_n(r,q)  \,,
\label{ms17}
\end{equation}
where the radial function  $\Re_n(r,q)$ is given by
$$
\Re_n(r,q) = \sqrt{\frac{R}{r}}\left\{ {\cal A}_{n}
\left[ \Gamma\left(n{+}\frac{1}{2} \right) \left(\frac{1}{2}qR\right)^{-\left(n+\frac{1}{2}\right)}\right] J_{n+\frac{1}{2}}(qr)+
\right.
$$
\begin{equation}
\left.
+{\cal B}_{n} \left[(-1)^n \frac{\pi}{\Gamma\left(n{+}\frac{1}{2} \right)} \left(\frac{1}{2}qR \right)^{n+\frac{1}{2}}\right] J_{-\left(n+\frac{1}{2}\right)}(qr) \right\} \,,
\label{ms142}
\end{equation}
with the Bessel function of the first kind with the half-integer index  $J_{n+\frac{1}{2}}(qr)$, adjoint Legendre polynomials $P^{(m)}_{n}$, Gamma-functions $\Gamma(s)$, and integration constants ${\cal A}_{n}$, ${\cal B}_{n}$.
The main new feature is the following: the azimuthal component
$B_{({\rm azim})}$, being equal to zero at $q{=}0$, becomes
non-vanishing at $q\neq 0$. For instance, for the model with dipole-type terrestrial magnetic field we obtain
\begin{equation}
B_{({\rm rad})}(r,q) = - \frac{2 \mu}{r^3} \cos{\theta}
\left(\cos{qr} + qr \sin{qr}\right) \,, \label{ms159}
\end{equation}
\begin{equation}
B_{({\rm merid})}(r,q) =    \frac{\mu \sin{\theta}}{r^3}
\left[\left(\cos{qr} + qr \sin{qr}\right) - q^2 r^2 \cos{qr}
\right]
 \,,
\label{ms1699}
\end{equation}
\begin{equation}
B_{({\rm azim})}(r,q) =  - q  \sin{\theta}\frac{\mu}{r^2}
\left(\cos{qr} + qr \sin{qr}\right) \,. \label{ms917}
\end{equation}
If the photon-axion coupling is absent, $q{=}0$, (\ref{ms159})-(\ref{ms917}) give standard formulas for the static dipole geomagnetic field
\begin{equation}
B_{({\rm rad})}(r,0) = - \frac{2 \mu}{r^3} \cos{\theta} \,, \quad
B_{({\rm merid})}(r,0) =    \frac{\mu \sin{\theta}}{r^3}  \,,
\quad B_{({\rm azim})}(r,0) = 0 \,. \label{ms9179}
\end{equation}
Clearly, the relic DM axions  deform the static terrestrial
magnetic field: while the original geomagnetic field
has the radial and meridional components only, the axion-photon
coupling produces a supplementary azimuthal component; this effect contributes to the
phenomenon of the Earth's magnetic pole drift. Also, the
axion-photon coupling provides the dependence of the magnetic field on the
altitude to become non-monotonic (see (\ref{ms159})-(\ref{ms917})).

\subsubsection{Axionically Induced Longitudinal Magneto-Electric Oscillations}

When we deal with axionically coupled oscillations in the spherical resonator,
bounded by the Earth surface ($r{=}R$) and the bottom
edge of the Earth Ionosphere ($r{=}R_{*}$), we enter the world of standing and running waves with frequencies of wide range,
which are generated by various geophysical processes and human activity. We are interested to extract the information about axionically induced electromagnetic oscillations.
To obtain the solution of the corresponding electrodynamic problem we reduced equations (\ref{1model6}) to a pair of equations for two potentials, $U$ and $V$, the analogs of known Debye potentials (see \cite{BG2013}).
In more details, we consider the following representation of radial, meridional, azimuthal components of electric and magnetic fields:
\begin{equation}
F_{0r} = - \frac{1}{r \sin{\theta}} \frac{\partial}{\partial \theta}(V \sin{\theta}) = E_{({\rm rad})} \,,
\quad F_{\theta \varphi} = r \frac{\partial}{\partial \theta} (U \sin{\theta}) = - r^2 \sin{\theta} B_{({\rm rad})} \,,
\label{t61}
\end{equation}
\begin{equation}
F_{\theta 0} = -  \frac{\partial}{\partial r} (r V) = -r E_{({\rm merid})} \,, \quad F_{r \varphi} = \sin{\theta} \frac{\partial}{\partial r} (r U)= - r \sin{\theta} B_{({\rm merid})}
\,,
\label{t62}
\end{equation}
\begin{equation}
F_{\varphi 0} = -  r \sin{\theta} \frac{\partial}{\partial x^0} U = -  r \sin{\theta} E_{({\rm azim})}
\,, \quad F_{r \theta} = r \left(qU +  \frac{\partial}{\partial x^0} V \right)= -r B_{({\rm azim})} \,,
\label{t623}
\end{equation}
and decompose the Debye potentials according to the requirements of boundary value problem on the borders of spherical resonator:
\begin{equation}
U(\tilde{t},r,\theta) = \sum_{n=0}^{\infty} \sum_{j=0}^{\infty}
u_{nj}(\tilde{t}) \ P^{(1)}_n(\cos{\theta}) \ {\cal H}_{nj}(r) \,,
\label{eig6}
\end{equation}
\begin{equation}
V(\tilde{t},r,\theta) = \sum_{n=0}^{\infty} \sum_{j=0}^{\infty}
v_{nj}(\tilde{t}) \ P^{(1)}_n(\cos{\theta}) \ {\cal H}_{nj}(r) \,.
\label{eig7}
\end{equation}
Here the radial functions ${\cal H}_{nj}(r)$ are
\begin{equation}
{\cal H}_{nj}(r) {=} \frac{1}{\sqrt{r}}\left[
J_{n{+}\frac{1}{2}}\left(\nu^{(n)}_{j} r \right)
J_{{-}\left(n{+}\frac{1}{2}\right)}\left(\nu^{(n)}_{j} R \right)
{-} J_{n{+}\frac{1}{2}}\left(\nu^{(n)}_{j} R \right)
J_{{-}\left(n{+}\frac{1}{2}\right)}\left(\nu^{(n)}_{j} r \right)
\right] \,,
\label{eig71}
\end{equation}
and the parameters $\nu^{(n)}_j$ can be extracted from the equation
\begin{equation}
J_{n+\frac{1}{2}}\left(\nu^{(n)}_{j} R \right)  J_{-\left(n+\frac{1}{2}\right)}\left(\nu^{(n)}_{j} R_{*}\right)
= J_{n+\frac{1}{2}}\left(\nu^{(n)}_{j} R_{*}\right) J_{-\left(n+\frac{1}{2}\right)}\left(\nu^{(n)}_{j} R\right)
\,,
\label{eig5}
\end{equation}
where the index $j=0,1,2,...$ counts the positive zeros of the
equation  $(\ref{eig5})$.
Then we obtain the coupled pair of equations for the mode amplitudes $u_{nj}$ and $v_{nj}$:
\begin{equation}
\ddot{u}_{nj} + c^2 \left[\left(\nu^{(n)}_{j}\right)^2 - q^2 \right]u_{nj} - qc \dot{v}_{nj} = 0
\,,
\label{eig8}
\end{equation}
\begin{equation}
\ddot{v}_{nj} + c^2 \left(\nu^{(n)}_{j}\right)^2 v_{nj} + qc \dot{u}_{nj} = 0
\,.
\label{eig9}
\end{equation}
Clearly, when $q=0$, i.e., when the axions are absent, the modes are decoupled; thus, namely the axions provide the interactions between $U$ and $V$ modes of oscillations.
In \cite{BG2013} we presented a complete analysis of the oscillation modes, but here we display the results for one case only, as an illustration of a resonant situation.
In the resonance case, when $q {=}\nu^{(n_{*})}_{j_{*}}$, for the mode with the number $n^*$ and index $j_{*}$, the corresponding amplitudes
\begin{equation}
v_{*}(\tilde{t}) =
\frac{1}{2}\left[v_{*}(0){-}\frac{\dot{u}_{*}(0)}{q} \right] +
\frac{1}{2}\left[v_{*}(0){+}\frac{\dot{u}_{*}(0)}{q} \right]
\cos{\sqrt{2} q \tilde{t}} + \frac{\dot{v}_{*}(0)}{\sqrt{2} q}
\sin{\sqrt{2} q \tilde{t}} \,, \label{res1}
\end{equation}
$$
u_{*}(\tilde{t}) = \left[u_{*}(0){+}\frac{\dot{v}_{*}(0)}{2q}
\right] -
\frac{q\tilde{t}}{2}\left[v_{*}(0){-}\frac{\dot{u}_{*}(0)}{q}
\right] +
$$
\begin{equation}
+ \frac{1}{2\sqrt2}\left[v_{*}(0){+}\frac{\dot{u}_{*}(0)}{q}
\right]\sin{\sqrt{2} q\tilde{t}} - \frac{\dot{v}_{*}(0)}{2q}
\cos{\sqrt{2} q\tilde{t}} \,, \label{res2}
\end{equation}
oscillate with the axionic frequency $\omega_{{\rm A}}{=}\sqrt2 q$, and the $U$ potential grows linearly with time ($\tilde{t}=x^0$, see (\ref{metric1})).

Main results of the analysis given in \cite{BG2013} are the following.

\noindent
1) Relic axions produce oscillations of a new type in the resonator "Earth-Ionosphere". We indicated them as Longitudinal Magneto-Electric Oscillations, since they possess the
following specific feature: the axionically coupled electric and magnetic
fields are parallel to one another. When the axions are absent and $q{=}0$, there exist only transversal electromagnetic oscillations, usual for the Faraday - Maxwell version of electrodynamics.
Longitudinal Magneto-Electric Oscillations can be considered as a dynamic analog of a static axionically induced effect predicted by Wilczek in \cite{Wlczk} (axions produce radial electric field in the vicinity of a monopole with radial magnetic field).

\noindent
2) New "hybrid" frequencies of oscillations appear in the global resonator "Earth-Ionosphere" due to the axionic Dark Matter influence.

\noindent
3) Estimations of the effect for $\rho_{({\rm
DM})} \simeq 1.25 \ {\rm GeV} \cdot {\rm cm}^{-3}$ and $\frac{1}{\Psi_0}{=} \rho_{{\rm A} \gamma \gamma} \simeq 10^{-9} {\rm GeV}^{-1}$, give
the value $\nu_{({\rm Axion})} \simeq 10^{-5} {\rm Hz}$ for the effective frequency of Longitudinal Magneto-Electric Oscillations in the Earth Magnetosphere.

\subsection{Electromagnetic Response on the Action of Gravitational pp-Waves in an Axionic Environment}
\label{GW1}

The second example of analysis of exact solutions to the equations of the minimal axion electrodynamics is associated with the case, when the gradient four-vector $\nabla_i \phi$ is the null one, i.e., $g^{ik} \nabla_i \phi \nabla_k \phi =0$. It is typical for models with the pp-wave symmetry \cite{Exact}.
In the framework of such models we use the background space-time metric of the form
\begin{equation}
ds^{2} = 2du dv - L^{2} \left[e^{2
\beta}(dx^2)^2 + e^{-2\beta} (dx^3)^2 \right] \,,
\label{GWmetric}
\end{equation}
which describes the gravitational pp-wave of the first polarization (see, e.g., \cite{MTW}). Here $u {=} \frac{ct {-} x^1}{\sqrt{2}}$ is the retarded time,
$v {=} \frac{ct {+} x^1}{\sqrt{2}}$ is the advanced time, and $L(u)$, $\beta(u)$ are the functions of the retarded time only.
The front of incoming plane gravitational wave is characterized by $u=0$, and we assume that $L(0)=1$, $L^{\prime}(0)=0$, $\beta(0)=0$ and $F^{*}_{mn}F^{mn}(0)=0$. After discovery of the gravitational waves reported in \cite{GW}, we obtained a new impetus to consider new problems associated with the influence of the gravitational radiation on axionically active media.

In \cite{BWTN2014} we solved the master equations of axion electrodynamics using two assumptions. Our first ansatz is that the potential of the pseudoscalar field has the form
\begin{equation}
V(\phi^2) = \frac12 \left[\frac{m^2_{({\rm a})}}{\nu} + \nu (\phi^2-\phi^2_{*})\right]^2  \,.
\label{b1}
\end{equation}
The second ansatz concerns the initial data. We assume,  that at $u<0$, i.e., before the gravitational wave appearing, the master equation for the pseudoscalar field
\begin{equation}
 \nabla^k \nabla_k \phi +  \left[m^2_{({\rm a})} + \nu^2 (\phi^2-\phi^2_{*})\right] \phi = - \frac{1}{4\Psi^2_0} F^{*}_{mn}F^{mn}
 \label{axioneq}
\end{equation}
admits the constant solution $\Phi$, which satisfies two conditions
\begin{equation}
V(\Phi^2) = 0 \,, \quad
\left[\frac{d}{d\phi}V(\phi_{}^2)\right]_{|\phi=\Phi} = 0 \,.
\label{b11}
\end{equation}
Clearly, this constant solution is of the form
\begin{equation}
\Phi = \pm \sqrt{\phi^2_{*} - \frac{m^2_{({\rm a})}}{\nu^2}} = \phi(0)\,.
\label{b19}
\end{equation}
We assume that $|\phi_{*}| > \frac{m_{({\rm a})}}{|\nu|}$, thus this potential
has two symmetric (real) minima.
As an illustration, in \cite{BWTN2014} we considered  the case, for which the initial electric field was absent, and magnetic field was constant,
$B_1(u<0)=B_{||}$, $B_2(u<0)=B_{\bot}\cos{\Theta}$, $B_3(u<0)=B_{\bot}\sin{\Theta}$.
In the field of gravitational wave, i.e., at $u>0$, the exact solutions of the set of master
equations of the axion electrodynamics can be represented as follows.
First, the solution for the axion field is
\begin{equation}
\phi(u) =  \phi(0) - 2 \arctan{
\left[\frac{\sin{2\Theta} \ \sinh{\beta(u)}}{\cosh{\beta(u)} +
\cos{2\Theta} \sinh{\beta(u)}} \right]}  \,,
\label{set5}
\end{equation}
where
\begin{equation}
a(u) \equiv \frac{1}{\sqrt{\cosh{2 \beta(u)} + \cos{2\Theta}
\sinh{2\beta(u)}}} \,, \quad a(0) = 1 \,.
 \label{set12}
\end{equation}
Second, the longitudinal magnetic
field $B_{||}$ is not distorted.
The longitudinal electric field $E_{||}(u)$
is proportional to the value $B_{||}$:
\begin{equation}
E_{||}(u)  =   \frac{2B_{||}}{L^2} \arctan{ \left[\frac{\sin{2\Theta}
\ \sinh{\beta}}{\cosh{\beta} + \cos{2\Theta} \sinh{\beta}}
\right]} = - \frac{B_{||}}{L^2}[\phi(u)-\phi(0)]  \,.
 \label{phys7}
\end{equation}
Third, the transversal components of the magnetic field are distorted:
\begin{equation}
{\cal B}^2 = L e^{\beta} B^{(2)} \left[1 + X(u,v) + Z(u)\right]
 \,,  \quad
 {\cal B}^3 = L e^{-\beta} B^{(3)} \left[ 1 + X(u,v)  - Z(u)\right] \,.
 \label{phys2}
\end{equation}
The transversal components of the electric field are generated under the influence of axionic and gravitational wave fields:
\begin{equation}
 {\cal E}^2 = \frac{B^{(3)}}{L} e^{\beta}  \left[ - Y(u,v)  - Z(u)\right]
 \,, \quad
 {\cal E}^3 = \frac{B^{(2)}}{L} e^{- \beta} \left[ Y(u,v)  - Z(u)\right]
 \,.
 \label{phys4}
\end{equation}
The distortion functions are defined as
\begin{equation}
 X(u,v) = \frac{1}{2} \left[ a(u) -1 - va^{\prime}(u) \right] \,,
 \quad Y(u,v) = \frac{1}{2} \left[ a(u) -1 + va^{\prime}(u) \right]
 \,,
 \label{phys5}
\end{equation}
\begin{equation}
 Z(u) =  \frac{2 \Psi^2_0 L^2}{a(u) B^2_{\bot}\sin{2\Theta}}
\left[{\cal H}(\Phi) {+} \frac{(B_{||})^2}{L^4 \Psi^2_0} \right]
\arctan{ \left[\frac{\sin{2\Theta} \ \sinh{\beta}}{\cosh{\beta}
{+} \cos{2\Theta} \sinh{\beta}} \right]} \,,
 \label{phys6}
\end{equation}
\begin{equation}
{\cal H}(\Phi) \equiv  m^2_{({\rm A})} + \nu^2 \left[\Phi^2 +
\Phi \phi(0) + \phi^2(0) - \phi^2_* \right] \,,
 \label{set14}
\end{equation}
where the prime denotes the derivative with respect to retarded time. These formulas display the symptom of anomalous behavior of the electromagnetic response on the gravitational wave action in the environment of axionic Dark Matter.
Indeed, let us compare the limit $\lim_{B_{\bot}\to 0}\{F_{ik}(B_{\bot})\}$ and the value $\{F_{ik}(B_{\bot}{=}0) \}$. Clearly, when $\beta \neq 0$, the function $Z(u)$ contains the parameter $B_{\bot}$ in the denominator, so the mentioned limit is infinite, while the second quantity, i.e., the absent electromagnetic field, is equal to zero. When $\beta = 0$, one sees that $Z(u)=0$, and two mentioned limits coincide.

This model describes, in fact a new mechanism of axion-photon-graviton coupling, which is associated with anomalous behavior of the electromagnetic response.
In this mechanism the axionic Dark Matter plays a provocative role of a mediator-amplifier.
Clearly, the constant pseudoscalar (axion) field $\phi$ is hidden from the point of view of axion electrodynamics; but this degeneracy happens to be removed after the appearance of the gravitational pp-wave. Then, the activated axion field generates the electric field proportional to the value of the initial magnetic field, and deforms the initial magnetic field.
Concerning the magnitude of the described effect, the very optimistic value for the term
$\frac{\Psi^2_0 m^2_{({\rm A})}}{B^2_{\bot}}$ in the function $Z(u)$ is estimated to be
of the order $10^{20}$ for the terrestrial magnetic field, and of the order $10^{28}$ for the magnetized
interstellar medium (see \cite{BWTN2014} for details and extended analysis of the model). These estimations, given for the axionically mediated electromagnetic response, are much more optimistic than the estimations for electromagnetic response induced by pure gravitational wave, which deforms an initially constant magnetic field in vacuum (see, e.g., \cite{Boc,BL2001}).

\section{Model 2. Non-stationary Optical Activity Induced by the Axionic Dark Matter}

\subsection{Extension of the Axion Electrodynamics: Inertia Effects and Field Theory}

We use the term "inertia effects" in an wide sense, when the Lagrangian of a model depends on the velocity four-vector $U^i$. It is known, that when one deals with scalar, electromagnetic, gauge, etc. fields in a standard vacuum, one uses the taboo on the introduction of the velocity four-vector $U^i$ into the Lagrangian.  Thus, insertion of the velocity four-vector $U^i$ is a symptom of consideration of a new coupling.

\subsubsection{Susceptibility of Spatially Isotropic Moving Medium}

The most known example of such extension appeared in the electrodynamics of spatially isotropic continua \cite{ED1}, where the additional term
\begin{equation}
 L_{(21)}= \frac14 \chi^{ikmn}_{(21)} F_{ik} F_{mn}
\label{iner00}
\end{equation}
was introduced with
\begin{equation}
 \chi^{ikmn}_{(21)} = \frac12 \left(\frac{1}{\mu}{-}1 \right) \left[g^{im}g^{kn}{-}g^{in}g^{km} \right] {+} \frac12 \left(\varepsilon {-}\frac{1}{\mu}\right) \left[g^{im} U^kU^n {-} g^{in}U^kU^m {+}
 g^{kn}U^iU^m {-} g^{km}U^iU^n \right]
 \,. \label{iner1}
\end{equation}
Here the phenomenological parameters $\varepsilon$ and $\mu$ are the dielectric and magnetic permittivities, respectively. In vacuum,  $\varepsilon{=}1$ and $\mu{=}1$, so $ \chi^{ikmn}_{(21)} {=}0$. Let us note that the nomenclature (21) in the term $ \chi^{ikmn}_{(21)}$ means that it is the first example in the second model. In our context, the four-vector $U^i$ describes the velocity of the Dark Energy, thus the term (\ref{iner1}) relates to a specific inertia-type interaction between the electromagnetic field and Dark Energy.

The additional term (\ref{iner00}) gives the following total induction tensor $H^{ik}$:
\begin{equation}
 H^{ik} = \frac{1}{\mu} F^{ik} + \left(\varepsilon -\frac{1}{\mu}\right) U_m \left[F^{im} U^k - F^{km}U^i \right]
 \,, \label{iner2}
\end{equation}
providing the well-known formulas for the total permittivities:
\begin{equation}
 \varepsilon^{im} = \varepsilon \Delta^{im} \,, \quad \left(\mu^{-1}\right)^{im} = \frac{1}{\mu}\Delta^{im} \,, \quad \nu^{pq} = 0
 \,, \label{iner3}
\end{equation}
and the constitutive equations
\begin{equation}
{\cal D}^i = \varepsilon E^i \,, \quad
{\cal H}_m =  \frac{1}{\mu} B_m  \,. \label{iner33}
\end{equation}
The total stress-energy tensor of the electromagnetic field in such medium is derived in \cite{B2007GC}; it has the form
\begin{equation}
 T^{({\rm EM})}_{ik} =
\frac{1}{4} g^{kl} H^{mn}
F_{mn}  - \frac{1}{2} (H^{km} F^l_{ \ m} +  H^{lm} F^k_{ \ m} )
\,. \label{BPisotrop}
\end{equation}
This stress-energy tensor is symmetric (e.g., as the Abraham tensor) and traceless (e.g., as the Minkowski tensor) (one can find the detailed discussion concerning the Abraham-Minkowski controversy, e.g., in the review \cite{ED5}). In order to derive (\ref{BPisotrop}) we used the formulas for variation of the velocity four-vector with respect to metric
\begin{equation}
\delta U^i =  \frac{1}{4} \delta g^{pq} \left(U_p \delta^i_q  {+}
U_q \delta^i_p \right) \,, \quad
\delta U_i =  - \frac{1}{4} \delta g^{pq} \left(U_p g_{iq}  {+}
U_q g_{ip} \right) \,,
\label{gr59}
\end{equation}
(see \cite{B2007GC,B2007CQG} for details).

\subsubsection{Axionically Induced Spontaneous Magnetization of the Inertia-Type }

There are no terms linear in $F_{mn}$, which contain the four-vector $U^i$ only, however, using the gradient four-vector $\nabla_i \phi$, as an additional element, one can construct the new term
\begin{equation}
L_{(22)}= \frac12 \lambda_{(22)} F^{*}_{mn} \nabla^m \phi \ U^n \,.
\label{agr1}
\end{equation}
This term contributes into the tensor of spontaneous polarization-magnetization
\begin{equation}
{\cal H}^{ik}_{(22)} = \frac12 \lambda_{(22)}\epsilon^{ikmn} \nabla_m \phi \ U_n \,,
\label{agr2}
\end{equation}
gives vanishing polarization, ${\cal P}^i=0$, but forms a spontaneous magnetization
\begin{equation}
{\cal M}_p = -\frac12 \lambda_{(22)} \nablab_p \phi \,.
\label{agr3}
\end{equation}
The corresponding contribution into the total stress-energy tensor is linear in the dual Maxwell tensor
\begin{equation}
T^{(22)}_{ik} = - \frac14 \lambda_{(22)} \nabla^m \phi \left(U_i F^{*}_{km}+ U_k F^{*}_{im}\right) \,.
\label{agr4}
\end{equation}
Finally, let us mention that the term (\ref{agr1}) changes the master equation for the axion field by adding the source
\begin{equation}
{\cal J}_{(22)} = \frac12 \lambda_{(22)} F^{*mn} \nabla_{[m} U_{n]}
\label{agr5}
\end{equation}
into the right-hand side of (\ref{phi1}).
As usual, the symbol  $\nabla_{[m} U_{n]}$ indicates the skew-symmetrization, $\nabla_{[m} U_{n]} \equiv \frac12 \left[\nabla_{m} U_{n} {-}\nabla_{n}U_{m}\right]$.
We repeat, that the nomenclature (22) indicates that we deal with term number two in the second model.

\subsubsection{Axionically Induced Optical Activity of the Inertia-Type}

Let us extend the Lagrangian  by the terms quadratic in the Maxwell tensor $F_{mn}$, linear in
$\phi$ or in $\nabla_k \phi$, and containing the velocity four-vector $U^k$. Clearly, in order to form irreducible scalar invariants of such type we
have to use the convolution $F^{ik} F^{*}_{kj}$. As it was shown in the Appendix A of \cite{BMZ2014A}, this convolution satisfies the relations
\begin{equation}
F^{ik} F^{*}_{kj} = \frac{1}{4}\delta^i_j \ F^{mn} F^{*}_{mn}  \,. \label{star0}
\end{equation}
One can check directly, that the set of new possible terms of mentioned type can be reduced to one irreducible invariant only, namely
\begin{equation}
L_{(23)} = \frac{1}{4} \lambda_{(23)}\ F^{mn} F^{*}_{mn} \ U^k \nabla_k \phi  \,, \label{star3}
\end{equation}
with one new coupling constant $\lambda_{(23)}$. The corresponding contribution into the induction tensor
\begin{equation}
H^{ik}_{(23)} =  \lambda_{(23)}  F^{*ik} {\cal D} \phi
\label{eld2k}
\end{equation}
provides the only total tensor of magneto-electric coefficients to be extended
\begin{equation}
\nu_{p}^{\ m} = - \Delta_p^m \left[\phi + \lambda_{(23)} {\cal D} \phi \right] \,. \label{eld79}
\end{equation}
The electrodynamic equations takes now the form
\begin{equation}
\nabla_k F^{ik} = - F^{*ik} \nabla_k [\phi + \lambda_{(23)} D\phi] \,.
\label{nu22}
\end{equation}
Equation for the axion field evolution obtains the following new term in the right-hand side:
\begin{equation}
{\cal J}_{(23)} =
\frac{\lambda_{(23)}}{4\Psi_{0}^2}\left(\Theta + D \right)\left(F^{*}_{mn}F^{mn}\right) \,.
\end{equation}
The backreaction of this coupling to the gravity field is described by the source-term
\begin{equation}
T_{pq}^{(23)} =  - \frac{1}{8} \lambda_{(23)} \ F^{mn}F^{*}_{mn} \left(U_p\nabla_q \phi {+}U_q \nabla_p \phi \right)
\label{gr42}
\end{equation}
in the right-hand side of the gravity field equations.

\subsection{An Illustration}

When a test electromagnetic wave
coupled to the axionic Dark Matter propagates in the spatially homogeneous
FLRW-type space-time with the scale factor $a(t)$ (say, in the direction $0x$),
in the short wavelengths approximation $k>>H(t)$
we obtain the solution for circularly polarized electromagnetic wave (see \cite{BT2012}) in the form
\begin{equation}
A_2(t,x) = - A_0 \sin{\left[W - \varphi(t)\right]} \,, \quad A_3(t,x) = A_0
\cos{\left[W - \varphi(t) \right]} \,, \quad A_2^2+A_3^2=A_0^2 \,. \label{S6}
\end{equation}
Here the phase  of the wave is given by the function
\begin{equation}
 W = W(t_0) + k\left[\int_{t_0}^t \frac{dt'}{a(t')} - x \right] \,, \quad \varphi(t) \equiv \Phi(t)-\Phi(t_0)
\,, \quad \Phi(t) = \frac12 [ \phi(t) + \lambda_{(23)} \dot{\phi}] \,.\label{S21}
\end{equation}
When $\lambda_{(23)}=0$, we deal with the well-known axionically induced polarization rotation: the angle of the phase shift $\varphi$ and the axion field $\phi$ differ by the coefficient $\frac12$ (see, e.g., \cite{Ni2,Ni3}).
When $\lambda_{(23)}\neq 0$, the angle of the polarization rotation depends on $\dot{\phi}$, i.e., on the rate of pseudoscalar field evolution.
Let us remind that in the cosmological context the function $\dot{\phi}$ can be
represented in terms of the Dark Matter energy-density $W^{({\rm DM})}$
and pressure $P^{({\rm DM})}$ according to (\ref{bili11}).
Thus, the extended axion electrodynamics can be considered as a
tool for investigation of the nonstationary effects in the
evolution of the axionic Dark Matter, caused by a retardation of the response, or in other words, caused by interactions of rheological type.

\section{Model 3. Gradient-type Interactions with the Axionic Dark Matter}

\subsection{Extended  Axion Electrodynamics: Taking into Account Terms Quadratic in the Gradient Four-Vector}

When we speak about quadratic terms in the gradient four-vector $\nabla_k \phi$, we do not use the velocity four-vector $U^i$ in the Lagrangian, and keep in mind, that there are only two irreducible terms of this type to be included into the Lagrangian (see \cite{BBT2012}):
\begin{equation}
L_{(3)} = \frac{1}{4}\lambda_{(31)} F_{mn}F^{mn}  \ \nabla^p \phi \nabla_p \phi {+} \frac{1}{4} \lambda_{(32)} F_{mp} F^{mq}  \ \nabla^p \phi \nabla_q \phi  \,,
\label{quad1}
\end{equation}
where the parameters $\lambda_{(31)}$ and $\lambda_{(32)}$ are phenomenological coupling constants.
Additional term appeared in the induction tensor is now of the form
\begin{equation}
H^{ik}_{(3)} =  \lambda_{(31)}  F^{ik}  \ \nabla_q \phi \nabla^q \phi
+ \lambda_{(32)} \nabla^{[k} \phi F^{i]q} \nabla_q \phi \,.
\label{eld29}
\end{equation}
The corresponding term in the susceptibility tensor reads
$$
\chi^{ikmn}_{(3)} = \frac{1}{2} \left( g^{im} g^{kn} {-} g^{in} g^{km} \right) \lambda_{(31)} \nabla_p \phi \nabla^p \phi {+}
$$
\begin{equation}
{+}
\frac{1}{2} \lambda_{(32)} \left( g^{i[m} \nabla^{n]} \phi \nabla^k \phi
{+} g^{k[n} \nabla^{m]} \phi \nabla^i \phi \right) \,.
\label{eld4}
\end{equation}
The total tensors
$\varepsilon^{im}$, $(\mu^{-1})_{im}$ and $\nu^{pm}$ take now the form
\begin{equation}
\varepsilon^{im} = \Delta^{im}\left[1{+} \lambda_{(31)} \nabla_q \phi \nabla^q \phi \right] +
\frac{1}{2}\lambda_{(32)} \left[ \Delta^{im} \left(  D \phi \right)^2 {+} \nablab^i \phi \nablab^m \phi \right]\,,
\label{eld11}
\end{equation}
\begin{equation}
\left(\mu^{-1}\right)_{im} = \Delta_{im} \left[ 1+ \lambda_{(31)} \nabla_q \phi \nabla^q \phi \right]+
\frac{1}{2}\lambda_{(32)} \left[\Delta_{im}\nablab_q \phi \nablab^q \phi - \nablab_i \phi \nablab_m \phi \right] \,,
\label{eld12}
\end {equation}
\begin{equation}
\nu^{pm} = - \phi \Delta^{pm} + \frac{1}{2} \lambda_{(32)}  D\phi \ \eta^{pmk}
\nablab_k \phi \,.
\label{eld13}
\end{equation}
The equation of evolution of the pseudoscalar field contains  additional terms in the left-hand side
\begin{equation}
\nabla_q \left[ \left(g^{pq}  {-} \Theta^{pq}\right) \nabla_p \phi \right] {+} \phi V' \left( \phi^2 \right)=
{-} \frac{1}{4 \Psi_{0}^2 } F^{*}_{mn} F^{mn} \,.
\label{ps1}
\end{equation}
Here the tensor $\Theta^{pq}$ is introduced as follows:
\begin{equation}
\Theta^{pq} = \frac{1}{2\Psi_{0}^2 } \left[\lambda_{(31)} g^{pq} F_{mn} F^{mn} {+} \lambda_{(32)} F^{mp} F_{m}^{\ q} \right]\,.
\label{ps2}
\end{equation}
The principal novelty of the model is that the tensor
\begin{equation}
\tilde{g}^{pq} = \left(g^{pq}  - \Theta^{pq}\right)
\label{ps3}
\end{equation}
plays the role of an effective metric for the axionic waves in analogy with color and color-acoustic metrics studied in  \cite{BZD1,BZD2,BZD3}.
Gravity field equations are extended by the interaction source-term
\begin{equation}
T_{pq}^{(3)} = \lambda_{(31)} T_{pq}^{(31)} + \lambda_{(32)}T_{pq}^{(32)} \,,
\label{gr49}
\end {equation}
where the tensors
\begin{equation}
T_{pq}^{(31)} = - \frac{1}{2} F_{mn}F^{mn} \nabla_p \phi \nabla_q \phi + T_{pq}^{(EM)} \nabla_n \phi \nabla^n
\phi \,,
\label{gr4}
\end {equation}
\begin{equation}
T_{pq}^{(32)} = \frac{1}{4} g_{pq} F_{m}^{\ n} F^{ml} \nabla_n \phi \nabla_l \phi {-} \frac{1}{2}\nabla^l \phi \left[F_{ql} F_{p}^{\ m}
  \nabla_m \phi  {+} F_{ml} F^{m}_{\ q}  \nabla_p \phi {+} F_{ml} F^{m}_{\ p}  \nabla_q \phi \right] \,,
\label{gr5}
\end{equation}
are quadratic both in the Maxwell tensor $F_{mn}$ and in the gradient four-vector $\nabla_k \phi$.

\subsection{First Illustration: a Spatially Homogeneous Anisotropic Cosmological Model}

Let us consider the Bianchi-I model with magnetic field. This model with the metric \cite{Exact}
\begin{equation}
ds^2 = dt^2 {-} a^2(t) \ (dx^1)^2 {-} b^2(t) \ (dx^2)^2 {-} c^2(t) \
(dx^3)^2 \,, \label{metric}
\end{equation}
was used in hundreds of works for various cosmological contexts. (Keeping in mind applications to the DF electrodynamics, we quote here only four papers \cite{MF1,MF2,MF3,MF4}).  We assume that all the state functions depend on the cosmological time only.
Direct calculations based on (\ref{eld11})-(\ref{eld13}) show that for the Model 3 with such space-time the dielectric permittivity tensor contains both new coupling constants
\begin{equation}
\varepsilon^{im} =  \varepsilon(t) \ \Delta^{im}\,, \quad \varepsilon(t)= 1 + \left[\lambda_{(31)}+ \frac{1}{2}\lambda_{(32)} \right] \dot{\phi}^2 \,,
\label{ueld11}
\end{equation}
the magnetic impermeability tensor includes only one new coupling constant, $\lambda_{(31)}$,
\begin{equation}
\left(\mu^{-1}\right)_{im} =  \frac{1}{\mu(t)} \Delta_{im} \,, \quad
\frac{1}{\mu(t)} = 1{+} \lambda_{(31)} \dot{\phi}^2 \,,
\label{ueld12}
\end {equation}
and the magneto-electric cross-effect tensor contains neither $\lambda_{(31)}$, nor $\lambda_{(32)}$.
The square of the refraction index
\begin{equation}
n^2(t) = \varepsilon(t) \mu(t) = \frac{1 + \left[\lambda_{(31)}+ \frac{1}{2}\lambda_{(32)} \right] \dot{\phi}^2}{1{+} \lambda_{(31)} \dot{\phi}^2}
\label{ueld139}
\end{equation}
depends now on cosmological time through the function ${\dot{\phi}}^2(t)$.
As it was shown explicitly in \cite{BBT2012} the interaction of axionic Dark Matter with global magnetic field generates an electric field $E^3{=} F^{30}(t)$ parallel to the magnetic field $B^{3}(t) {=} {-} \frac{1}{abc}F_{12}$ (it is the typical axionically induced Longitudinal Magneto-Electric Cluster). For an illustration, we display here the exact solution for the electric field, which satisfies the condition  $F^{30}(t_0)=0$ at some moment of the cosmological time $t_0$
\begin{equation}
F^{30}(t) {=} \frac{F_{12}[ \phi(t) -\phi(t_0)]}{a(t)b(t)c(t)\left[1 {+} \left(\lambda_{(31)} {+} \frac{1}{2} \lambda_{(32)}\right) \dot{\phi}^2(t) \right]}
\,.
\label{cosm301}
\end{equation}
Clearly, when the coupling constants vanish, i.e., $\lambda_{(31)}{=}\lambda_{(32)}{=}0$, we deal with standard vacuum with $n^2{=}1$ and without anomalies in the electromagnetic field.
For non-vanishing coupling constants there are six intrinsic cases.

\vspace{3mm}
\noindent
{\it (i)} When  $\lambda_{(31)}{+}\frac{1}{2}\lambda_{(32)} \geq 0$ and $\lambda_{(31)}\geq 0$, there is no anomaly in the electric field, and the quantity $n^2(t)$ is always positive.

\vspace{3mm}
\noindent
{\it (ii)} When  $\lambda_{(31)}{+}\frac{1}{2}\lambda_{(32)} \geq 0$, and $\lambda_{(31)}< 0$, there is no anomaly in the electric field, but
the quantity $n^2(t)$ can take infinite value at some moment $t^{*}$, for which $ \dot{\phi}^2(t^{*})= \frac{1}{|\lambda_{(31)}|}$.
For infinite refraction index, the phase velocity of electromagnetic waves $V_{({\rm ph})}{=}\frac{1}{n}$ and the group velocity $V_{({\rm gr})}{=}\frac{2n}{n^2{+}1}$ take zero values, thus, the electromagnetic energy-information transfer stops. During the interval of cosmological time, for which  $ \dot{\phi}^2(t)> \frac{1}{|\lambda_{(31)}|}$ the square of refraction index is negative.
Such a situation is indicated in \cite{BBL2012} as unlighted epoch in the Universe history, since electromagnetic waves can not propagate in the Universe, when $n$ is pure imaginary quantity. Also, one can say, that it can be called a Dark Epoch of the first kind provided by the coupling of photons to the Dark Matter.

\vspace{3mm}
\noindent
{\it (iii)} When  $\lambda_{(31)}{+}\frac{1}{2}\lambda_{(32)} < 0$ and $\lambda_{(31)}\geq 0$,  a dynamic anomaly in the electric field can appear, if the time moment $t^{**}$ exists, for which  $ \dot{\phi}^2(t^{**})= \frac{1}{|\lambda_{(31)}+ \frac{1}{2}\lambda_{(32)}|}$. The quantity $n^2(t)$ can change the sign at $t^{**}$ providing the existence of a Dark Epoch of the second kind. On the boundary of this Epoch $n^2(t^{**}){=}0$, $V_{({\rm ph})}(t^{**}){=}\infty$, and the group velocity $V_{({\rm gr})}(t^{**})=0$, i.e., the electromagnetic energy transfer stops.

\vspace{3mm}
\noindent
{\it (iv)} When  $\lambda_{(31)}{+}\frac{1}{2}\lambda_{(32)} < 0$, $\lambda_{(31)}< 0$, and $0<\frac{1}{2}\lambda_{(32)} < |\lambda_{(31)}|$,
again a dynamic anomaly in the electric field can appear, and the quantity $n^2(t)$ can be negative, when
\begin{equation}
\frac{1}{|\lambda_{(31)}|}< {\dot{\phi}}^2<\frac{1}{|\frac{1}{2}\lambda_{(32)}-|\lambda_{(31)}||} \,.
\label{qq1}
\end{equation}
On the boundary of the corresponding Dark Epoch $n^2(t^{*}){=}\infty$, $V_{({\rm ph})}(t^{*}){=}0$, $V_{({\rm gr})}(t^{*})=0$.

\vspace{3mm}
\noindent
{\it (v)} When  $\lambda_{(31)}{+}\frac{1}{2}\lambda_{(32)} < 0$, $\lambda_{(31)}> \frac12 |\lambda_{(32)}|$, and
${\dot{\phi}}^2>\frac{1}{|\frac{1}{2}\lambda_{(32)}+\lambda_{(31)}|} $
a dynamic anomaly in the electric field can appear, and the quantity $n^2(t)$ also can be negative.
On the boundary of the corresponding Dark Epoch $n^2(t^{**}){=}0$, $V_{({\rm ph})}(t^{**}){=}\infty$, $V_{({\rm gr})}(t^{**})=0$.

\vspace{3mm}
\noindent
{\it (vi)} When  $\lambda_{(32)} = 0$, but $\lambda_{(31)} \neq 0$, one obtains that  $n^2=1$, however, $\varepsilon \neq 1$ and $\mu \neq 1$. There are no Dark Epochs, nevertheless, the anomaly in the electric field can exist, if $\lambda_{(31)}$ is negative and $\dot{\phi}^2(t)> \frac{1}{|\lambda_{(31)}|}$ for some interval of the cosmological time.

Let us emphasize that  according to (\ref{bili11}) one can replace ${\dot{\phi}}^2$ with $ {\dot{\phi}}^2 = \frac{1}{\Psi^2_0} \left[W^{({\rm DM})} {+} P^{({\rm DM})} \right]$, so in the framework of this model, the inequalities discussed above in the context of Dark Epochs, involve the state functions of the Dark Matter. In this context, some combinations of coupling constants can be expressed via the critical values of the DE mass density. For instance, let the Dark Matter be cold, and the coupling constants satisfy the inequalities related to the item {\it (v)}. Then the inequality (\ref{qq1}) takes the form $\rho_{(1)}< \rho_{({\rm DM})}<\rho_{(2)}$, where two critical values of the DE mass density, $\rho_{(1)}\equiv \frac{\Psi^2_0}{|\lambda_{(31)}|}$, and, $\rho_{(2)} \equiv \frac{\Psi^2_0}{|\frac{1}{2}\lambda_{(32)}-|\lambda_{(31)}||}$, are introduced.

\subsection{Second Illustration: Static Model with Spherical Symmetry}
\label{subsubsec252}

We consider the metric of static spherically symmetric field configurations to be of the form
\begin{equation}
ds^2 = \sigma^2(r) N(r) dt^2 {-} \frac{1}{N(r)} dr^2 - r^2 \left(d\theta^2 + \sin^2{\theta}d\varphi^2 \right) \,, \label{metric2}
\end{equation}
and assume that $\phi$ also depends on $r$ only. Since $\dot{\phi}=0$, we obtain that $I^i=0$. The energy density, longitudinal and transversal pressures of the Dark Matter are linked now by the following relationships
\begin{equation}
W^{({\rm DM})}=\frac{1}{2}\Psi^2_0 \left[V(\phi^2)+N \phi^{\prime 2} \right]
\,, \label{sss1}
\end{equation}
\begin{equation}
-P^{({\rm DM})}_{\bot} \equiv {\cal P}^{\theta}_{\theta} = {\cal P}^{\varphi}_{\varphi} = W^{({\rm DM})} \,, \label{sss22}
\end{equation}
\begin{equation}
-P^{({\rm DM})}_{||} \equiv {\cal P}^{r}_{r} = \frac{1}{2}\Psi^2_0 \left[V(\phi^2)- N \phi^{\prime 2} \right] \,. \label{sss2}
\end{equation}
The prime denotes the derivative with respect to $r$. Thus the pseudoscalar field $\phi(r)$ can be reconstructed using formula
\begin{equation}
\phi^{\prime }(r) = \pm \frac{1}{\Psi_0 \sqrt{N}} \sqrt{W^{({\rm DM})}+P^{({\rm DM})}_{||}} \,. \label{sss3}
\end{equation}
As an illustration, we consider  the solution for the model with monopole, which possesses a Longitudinal Magneto-Electric Cluster formed by collinear radial magnetic field and radial electric field induced by the axion-photon coupling. This solution is asymptotically flat with $\phi(\infty)=0$; it has the form
\begin{equation}
F_{\theta \varphi} = \mu \sin{\theta} \,, \quad F^{0r}(r)= \frac{\mu \phi(r)}{\sigma r^2 \left[1{-} N\phi^{\prime 2}\left(\lambda_{(31)} {+} \frac{1}{2}\lambda_{(32)}\right)\right]} \,. \label{sss4}
\end{equation}
When $\lambda_{(31)} {+} \frac{1}{2}\lambda_{(32)} \leq 0$, the axionically induced electric field is regular. When $\lambda_{(31)} {+} \frac{1}{2}\lambda_{(32)}> 0$, the spatial anomaly can appear at $r=r^{*}$, where $r^{*}$ satisfies the equation
$\phi^{\prime 2}(r^{*}) {=} N^{-1}(r^{*}) [\left(\lambda_{(31)} {+} \frac{1}{2}\lambda_{(32)}\right)]^{-1}$. Let us emphasize that according to the formula (\ref{sss3}) the quantity $\phi^{\prime 2}$
can be expressed in terms of state functions of the axionic Dark Matter distributed in the vicinity of spherically symmetric monopole.

\section{Model 4. Dynamo-Optical Interactions Associated with the Axionic Dark Matter}

In classical electrodynamics of continuous media there exists the term {\it dynamo-optical phenomena}, which  describes electromagnetic effects caused by a non-uniform motion of the medium \cite{ED2}.  Mathematically, these effects can be described by introduction into the Lagrangian terms linear in the covariant derivative of the macroscopic velocity four-vector $\nabla_iU_k$, or equivalently, the terms including its irreducible elements, the acceleration four-vector $DU^i$, shear tensor $\sigma^{ik}$, vorticity tensor $\omega^{ik}$, and expansion scalar $\Theta$ \cite{AB2006}. In this Section we deal with the Lagrangian, which includes the Maxwell tensor $F_{ik}$, the velocity four-vector $U^i$ and its covariant derivative $\nabla_i U_k$, as well as, the axion field $\phi$ and its gradient four-vector $\nabla_i \phi$. The presence of $\phi$  or $\nabla_i \phi$ in the Lagrangian $L_{(4)}$ allows us to link this model with dynamo-optical phenomena induced by the axionic Dark Matter.

\subsection{Axionic Extension of the Theory of Dynamo-Optically Active Electrodynamic Systems: The Lagrangian}

In order to list all irreducible dynamo-optical terms we have proposed in \cite{BA2014} the following strategy: in the decomposition of the Lagrangian instead of tensor $F_{ik}$ we used the four-vector of electric field $E^i$ and (pseudo) four-vector of magnetic induction $B^k$ (see (\ref{EB0}); instead of the tensor $\nabla_iU_k$ we used its representation (\ref{act3})-(\ref{act4}); instead of gradient (pseudo) four-vector $\nabla_i \phi$ we used the convective derivative $D\phi$ and spatial gradient $\nablab_i \phi$ orthogonal to $U^i$, taken from the decomposition
$\nabla_i \phi=U_i D\phi {+} \nablab_i \phi$.
The corresponding terms are presented by using the following nomenclature.
$$
L_{(4)} =  \frac{1}{4}E_m B_n \left[\phi+\lambda_{(40)} D\phi \right]\left[\lambda_{(41)} \Theta g^{mn}
{+} \lambda_{(42)}\sigma^{mn} {+} \lambda_{(43)}\omega^{mn} \right] +
$$
$$
+ \frac{1}{4} \omega_{(m} \nablab_{n)} \phi \left[
\Delta^{mn}\lambda_{(44)} \left(E_k E^k -B_k B^k \right) +  \lambda_{(45)} \left( E^m E^n + B^m B^n\right) \right]+
$$
$$
+  \frac{1}{4} \nablab_{n}\phi \ DU_{m}  \left[ \lambda_{(46)}
\Delta^{mn} E^k B_k +  \lambda_{(47)} \left( E^n B^m +
 E^m B^n \right) \right]+
$$
\begin{equation}
+ \frac{1}{4} \lambda_{(48)}  \eta^{nmp} \sigma_{mk}\nablab_{n}\phi
\left(E^k E_p {+} B^k B_p \right) \,. \label{decompSO9}
\end{equation}
Terms in the first line of this decomposition are linear in the pseudoscalar field $\phi$ and in the convective derivative $D\phi$; they contain pseudovector $B^i$
to provide the invariant property of this part of the Lagrangian. Other lines contain terms linear in the spatial gradient of the pseudoscalar
field, $\nablab_k \phi$. The second line in (\ref{decompSO9}) contains scalars linear in the vorticity tensor with $\omega_i {=} {-} \eta_{ikl}\omega^{kl}$ , thus guaranteeing the product
$\omega_{m} \nablab_{n} \phi $ to be the pure tensor. The third line in (\ref{decompSO9}) includes the terms linear in the acceleration four-vector
$DU_k$ and linear in $B_k$. The fourth line is formed with the terms linear in the shear tensor $\sigma^{ik}$, the term $\eta^{nmp} \nablab_{n}\phi$ being the pure tensor.

\subsection{Susceptibility of the Axionically Active Dynamo-Optical Medium}

In the context of this Section, in order to calculate the new contributions into the induction tensor, it is convenient to use the formula
\begin{equation}
H^{ik}_{(4)} = 2U^n \left[\delta^{ik}_{mn} \frac{\partial }{\partial E_m} + \epsilon^{ik}_{\ \ mn} \frac{\partial
}{\partial B_m} \right]L_{(4)} \,, \label{ED2}
\end{equation}
then to use (\ref{EB}) and (\ref{linlawDH}) to find new contributions into the permittivity tensors. Direct analysis yields
\begin{equation}
\varepsilon^{im}_{(4)} = \lambda_{(44)} \Delta^{im} \left(\omega^n \nablab_n \phi \right) + \lambda_{(45)} \Delta^{j(i}\Delta^{m)n} \omega_{(j} \nablab_{n)} \phi
+ \lambda_{(48)} \nablab_n \phi \ \eta^{nj(i} \sigma^{m)}_j \,, \label{e4}
\end{equation}
\begin{equation}
\left(\mu^{-1}\right)^{im}_{(4)} = \lambda_{(44)} \Delta^{im} \left(\omega^n \nablab_n \phi \right) - \lambda_{(45)} \Delta^{j(i}\Delta^{m)n} \omega_{(j} \nablab_{n)} \phi
- \lambda_{(48)} \nablab_n \phi \ \eta^{nj(i} \sigma^{m)}_j \,, \label{m4}
\end{equation}
\begin{equation}
\nu^{im}_{(4)} = {-}\frac12 \left[\phi {+} \lambda_{(40)} D\phi \right]\left[\lambda_{(41)}\Delta^{im}\Theta {+} \lambda_{(42)}\sigma^{im} {-} \lambda_{(43)}\omega^{im} \right]
{-} \lambda_{(46)}\Delta^{im} DU^n \nablab_n \phi {-} \lambda_{(47)} DU^{(i} \nablab^{m)} \phi \,. \label{n4}
\end{equation}
Clearly, three coupling parameters  $\lambda_{(44)}$, $\lambda_{(45)}$, and $\lambda_{(48)}$ are included into the symmetric (true) tensors of dielectric permittivity $\varepsilon^{im}_{(4)}$ and magnetic impermeability $\left(\mu^{-1}\right)^{im}_{(4)}$. Six coupling parameters $\lambda_{(40)}$, $\lambda_{(41)}$, $\lambda_{(42)}$, $\lambda_{(43)}$, $\lambda_{(46)}$ and $\lambda_{(47)}$ enter the non-symmetric pseudo-tensor of magneto-electric cross-effect $\nu^{im}_{(4)}$.

\subsection{An Illustration}

When the space-time is spatially isotropic and homogeneous, and the metric is of the Friedmann type, we know that the velocity four-vector can be chosen as $U^i=\delta^i_0$, and
\begin{equation}
DU^i =0 \,, \quad \sigma_{ik}=0 \,, \quad \omega_{ik} = 0 \,, \quad \Theta = 3 H(t)\,.
\end{equation}
This means that the axionically induced couplings of the dynamo-optical type do not disturb the permittivities, i.e., $\varepsilon^{im}_{(4)} {=} 0$ and $\left(\mu^{-1}\right)^{im}_{(4)}=0$ (see (\ref{e4}), (\ref{m4})). However, the magneto-electric effects are generated, since
\begin{equation}
\nu^{im}_{(4)} = \frac32 \lambda_{(41)} H(t) \left[\phi(t) + \lambda_{(40)} \dot{\phi} \right] \Delta^{im} \neq 0\,. \label{n94}
\end{equation}
Again, as in the Model 2, we deal with polarization rotation, when electromagnetic waves propagate in the Universe. The phase of rotation in that model was
$\Phi_{(23)} = \frac12 \left[\phi {+} \lambda_{(23)} \dot {\phi}\right]$. Now this function has the form
\begin{equation}
\Phi_{(4)} = \frac32 \lambda_{(41)} H(t) \left[\phi + \lambda_{(40)} \dot{\phi} \right]\,.
\label{n69}
\end{equation}
Thus, in the model with additional Lagrangian $L_{(4)}$, the effect of axionically induced optical activity is mediated by dynamo-optical interactions, which is displayed in the rotation
function $\Phi_{(4)}$ via the multiplier $\Theta = 3 H(t)$, where $H$ is the Hubble function. The following new source term
\begin{equation}
{\cal J}_{(4)} = - \frac{\lambda_{(41)}}{4\Psi^2_0} \left\{(E^kB_k)\left[\Theta - \lambda_{(40)} \left( \dot{\Theta} + \Theta^2 \right)\right] - \lambda_{(40)} \Theta D (E^kB_k) \right\}
\label{n695}
\end{equation}
appears in the right-hand side of the evolutionary equation (\ref{phi1}) for the pseudoscalar (axion) field.

\section{Model 5: Striction-type Coupling via a Scalar Dark Energy}

\subsection{A Prologue}

The idea of interaction between electromagnetic field and scalar field $\psi$ (dilaton) was realized in the Field Theory  at the same time as the idea of coupling between electromagnetic field and pseudoscalar field $\phi$ (axion). There is a lot of papers (see, e.g., \cite{D1,D2,D3} for details and references), in which the Lagrangian of electromagnetic field was extended as
\begin{equation}
\frac14 F^{mn} F_{mn}  \ \rightarrow \  \frac14 \left[1 +{\cal K}(\psi)\right] F^{mn} F_{mn}  \,, \label{dilaton1}
\end{equation}
where ${\cal K}(\psi)$ is a scalar multiplier depending on the dilaton field and satisfying the condition ${\cal K}(0)=0$. Clearly, when the multiplier is a linear function of $\psi$, i.e., ${\cal K}(\psi)= \omega_0 \psi$ (see, e.g., the work of Bekenstein \cite {D1} concerning variations of the fine-structure constant), the additional term
$\frac14 \omega_0 \psi F^{mn} F_{mn}$ is the direct analog of the term $\frac14 \phi F^{mn} F^{*}_{mn}$, which is basic for the axion electrodynamics.
Electrodynamic models with the Lagrangian (\ref{dilaton1}) can be characterized by the induction tensor
$H^{ik} = \left[1 +{\cal K}(\psi)\right] F^{ik}$, thus these models correspond to the susceptibility tensor
\begin{equation}
\chi^{ikmn}(\psi) = \frac12 {\cal K}(\psi)\left(g^{im}g^{kn}-g^{in}g^{km} \right)  \,, \label{dilaton2}
\end{equation}
when we consider the action of the dilaton field on the electromagnetic filed as an influence of some effective medium with $\varepsilon =\frac{1}{\mu}= {\cal K}(\psi)$.
When we deal with scalar $\Psi$-representation of the Dark Energy, there is a simple way to extend this version of the Maxwell-dilaton theory by introduction of the gradient four-vector $\nabla_i \Psi$, of the velocity four-vector $U^i$ and its covariant derivative, into the extended Lagrangian. However, we do not intend to do it, since this way contains reasoning very similar to the ones used for pseudoscalar (axion) field $\phi$, and it would be simple repetition of calculations given above. Below we use the medium representation of the Dark Energy, and in order to interpret the result, we address to analogies from the classical electrodynamics of continuous media.

\subsection{Electro-Striction and Magneto-Striction Induced by a Dark Energy}

\subsubsection{Extension of the Susceptibility Tensor}

Let us consider the Dark Energy to be presented by the stress-energy tensor (\ref{Eigen3}), by its time-like eigen four-vector $U^i$ attributed to the velocity four-vector, and by  the pressure tensor ${\cal P}_{ik}$ (see (\ref{Eigen4})). Now we introduce a new term into the Lagrangian (see \cite{BD2014}):
\begin{equation}
L_{(5)} = \frac14 Q^{ikmnpq}   \ {\cal P}_{pq} \ F_{ik} F_{mn} \,, \label{stric1}
\end{equation}
linear in the pressure tensor and containing a six-indices tensor $ Q^{ikmnpq}$, components of which describe the electromagnetic response associated with
electro-striction and magneto-striction.
The corresponding contributions into the induction  and  susceptibility tensors are, respectively
\begin{equation}
H^{ik}_{(5)} =  Q^{ikmnpq}   \ {\cal P}_{pq} \ F_{mn} \,, \quad \chi^{ikmn}_{(5)} = Q^{ikmnpq}   \ {\cal P}_{pq} \,.
\label{stric3}
\end{equation}
The tensor $ Q^{ikmnpq}$ can be decomposed similarly to the linear response tensor (\ref{Cdecomposition}); for this purpose, we introduce the following four-indices tensors
$$
\alpha^{im(pq)} = 2Q^{ikmnpq} U_k U_n \,,
\quad \beta^{ls(pq)} = {-} \frac12 \eta^l_{\ ik} Q^{ikmnpq}\eta^s_{\ mn} \,,
$$
\begin{equation}
\quad \gamma^{lm(pq)} {=} \eta^l_{\ ik} Q^{ikmnpq}U_n \,.
\label{stric4}
\end{equation}
The convolutions of these tensors with the pressure ${\cal P}_{pq}$ give the striction-type contributions to the dielectric permittivity, $\varepsilon^{im}_{(5)}$, to the magnetic impermeability,
 $\left(\mu^{-1}\right)^{ls}_{(5)}$, and magneto-electric cross-effect tensor, $\nu^{lm}_{(5)}$ , respectively. The tensor $Q^{ikmnpq}$ possesses the following symmetries:
\begin{equation}
Q^{ikmnpq} = Q^{ikmnqp} = - Q^{kimnpq} = - Q^{iknmpq}= Q^{mnikpq} \,.
 \label{stric2}
\end{equation}
Thus, the tensors $\alpha^{im(pq)}$, $\beta^{ls(pq)}$, $\gamma^{lm(pq)}$ inherit the symmetry with respect to indices $(pq)$. Also,  $\alpha^{im(pq)}$ and $\beta^{ls(pq)}$ are symmetric with respect to indices in the first pair, but $\gamma^{lm(pq)}$ does not possess the last symmetry.

\subsubsection{Extension of the Gravity Field Equations}

The stress-energy tensor $T^{(5)}_{ik}$, calculated for the term $L_{(5)}$ can be written as follows:
$$
T^{(5)}_{ik} = \frac14 g_{ik} Q^{abmnpq}F_{ab}F_{mn} {\cal P}_{pq} -
\frac12 F_{ab}F_{mn} {\cal P}_{ls} \frac{\delta}{\delta g^{ik}}\left(Q^{abmnpq} \Delta^l_p \Delta^s_q \right)
+
$$
\begin{equation}
+ Q^{abmnpq}F_{ab} F_{mn} {\cal B}_{ikls}\Delta^l_p \Delta^s_q
\,, \label{S78}
\end{equation}
where the tensor ${\cal B}_{ikls}$ is introduced using the variation derivative of the second order
\begin{equation}
{\cal B}_{ikls} \equiv \frac{1}{\sqrt{-g}} \frac{\delta^2 }{
\delta g^{ik} \delta g^{ls}} \left[\sqrt{-g} \ L_{(5)}\right]
 \,. \label{S2}
\end{equation}
We will illustrate the calculations of the variation derivatives below in the application to the spatially isotropic model.

\subsection{Application to a Spatially Isotropic Homogeneous Dark Energy}

\subsubsection{Reduction of the Susceptibility Tensor}

When the electrodynamic system is spatially isotropic, we can put ${\cal P}_{ik}{=} -P \Delta_{ik}$ and decompose the space-like tensors
$\alpha^{im(pq)}$, $\beta^{im(pq)}$ and $\gamma^{im(pq)}$ using the metric, Kronecker deltas, Levi-Civita tensor and the velocity four-vector only.
These decompositions yield
\begin{equation}
\alpha^{im(pq)} {=} \alpha_{(1)} \Delta^{im} \Delta^{pq} {+}
\alpha_{(2)} (\Delta^{ip}\Delta^{mq}{+} \Delta^{iq}\Delta^{mp}) \,,
\label{abg1}
\end{equation}
\begin{equation}
\beta^{im(pq)} = \beta_{(1)} \Delta^{im} \Delta^{pq} + \beta_{(2)}
(\Delta^{ip}\Delta^{mq}+ \Delta^{iq}\Delta^{mp})\,, \quad \gamma^{im(pq)}=0 \,. \label{abg2}
\end{equation}
Thus, when the medium is
spatially isotropic, one deals with four independent coupling parameters
$\alpha_{(1)}$, $\alpha_{(2)}$, $\beta_{(1)}$, $\beta_{(2)}$. First two parameters characterize electro-striction induced by the Dark Energy,
the last two parameters relate to the DE-induced  magneto - striction. With these formulas we can reconstruct the tensor $Q^{ikmnpq}$ as follows:
$$
Q^{ikmnpq}= \frac12 \left[\alpha_{(1)}\Delta^{pq}
\left(g^{ikmn}{-}\Delta^{ikmn} \right) {+} \alpha_{(2)} U_l U_s
\left(g^{iklp} g^{mnsq}{+} g^{iklq}g^{mnsp}\right) + \right.
$$
\begin{equation}
\left. + \beta_{(1)}\Delta^{pq}\Delta^{ikmn} -
\beta_{(2)}(\eta^{ikp}\eta^{mnq}{+}\eta^{ikq}\eta^{mnp})
\right]\,. \label{abg4}
\end{equation}
Since ${\cal P}_{ik}{=} -P \Delta_{ik}$, only the four-indices tensor $Q^{ikmnpq} \Delta_{pq}$ appears in
the electrodynamic equations. Now we obtain that
\begin{equation}
Q^{ikmn} \equiv Q^{ikmnpq}\Delta_{pq} = \frac12 \alpha g^{ikmn} + \frac12 (\beta-\alpha)\Delta^{ikmn}
\,, \label{abg49}
\end{equation}
i.e., only two effective coupling constants
\begin{equation}
\alpha = 3 \alpha_{(1)} + 2
\alpha_{(2)} \,,  \quad \beta = 3
\beta_{(1)} + 2 \beta_{(2)} \,, \label{asm22}
\end{equation}
are essential.
Calculation of the total
permittivity tensors of the spatially isotropic striction-active medium yields
\begin{equation}
\varepsilon^{im} = \Delta^{im} \varepsilon \,, \quad \varepsilon =
\varepsilon_{(0)} - \alpha P\,, \label{e22}
\end{equation}
\begin{equation}
(\mu^{-1})_{ab} = \frac{1}{\mu} \ \Delta_{ab}  \,, \quad
\frac{1}{\mu} = \frac{1}{\mu_{(0)}} - \beta P \,, \quad \nu^{am} = 0 \,.
\label{m22}
\end{equation}
The square of the refraction index of such medium is
\begin{equation}
n^2 \equiv \varepsilon \mu =  \frac{n^2_{(0)} - \mu_{(0)} \alpha P}{1 - \mu_{(0)}\beta P } \,, \label{22varco}
\end{equation}
where $n^2_{(0)} \equiv \varepsilon_{(0)} \mu_{(0)}$.
With this refraction index we can find the phase and group velocities of the electromagnetic
waves in the striction-active medium
\begin{equation}
V_{({\rm ph})} \equiv \frac{1}{n}= \sqrt{\frac{1 - \mu_{(0)} \beta P}{n^2_{(0)} - \mu_{(0)} \alpha P}}\,, \quad
V_{({\rm gr})} \equiv \ \frac{2n}{({n}^2+1)} \,.
\label{v1}
\end{equation}
(Let us repeat that we use the system of units with $c=1$).

\subsubsection{Reduction of the Striction Source in the Gravity Field Equations}

For the spatially isotropic model the variation
derivative $\frac{\delta}{\delta g^{ik}}\left(Q^{abmnpq} \Delta^l_p
\Delta^s_q \right)$ can be calculated directly using the auxiliary formulas
\begin{equation}
 \frac{\delta \Delta^{pq}}{\delta g^{ik}} = \delta^{(p}_{(i} \Delta^{q)}_{k)} \,, \quad
\frac{\delta g^{abmn}}{\delta g^{ik}} = \delta^{[a}_{(i} g_{k)}^{\
b]mn} +  g_{\ \ \ \ (k}^{ab[m} \delta^{n]}_{i)} \,,
\label{1T2}
\end{equation}
\begin{equation}
 \frac{\delta \Delta^{abmn}}{\delta g^{ik}} =  \delta^{[a}_{(i} \Delta_{k)}^{\ b]mn} +  \Delta_{\ \ \ \ (k}^{ab[m} \delta^{n]}_{i)} \,, \quad
   \frac{\delta \eta^{abp}}{\delta g^{ik}} = \frac12 \left[\eta^{abp} g_{ik} - \epsilon^{abp}_{\ \ \ (i}U_{k)} \right] \,,
 \label{1T40}
\end{equation}
(see Appendix in \cite{BD2014}). As for the tensor ${\cal B}_{ikls}$, one can find the detailed calculations in \cite{BD2014}; it has the form
$$
{\cal B}_{ikls} = \frac14 \left(2 W {-} P \right)U_{i} U_{k} U_{l}
U_{s} +
\frac14 W \left(U_{i} U_{k} \Delta_{ls}+ U_{l} U_{s}
\Delta_{ik}\right) - P  U_{(l} \Delta_{s)(k}U_{i)} -
$$
\begin{equation}
 - \frac14 P \left(
\Delta_{ls}\Delta_{ik} {+} \Delta_{li}\Delta_{ks} {+}
\Delta_{lk}\Delta_{is}\right) \,. \label{DE7}
\end{equation}
With these two contributions one can reconstruct $T^{(5)}_{ik}$:
\begin{equation}
T^{(5)}_{ik} = Q^{abmn}
 \ F_{ab}\left\{ P \ \left[-\frac14 g_{ik}F_{mn}+\frac12
\left(g_{im}F_{kn}+g_{km}F_{in} \right) \right]  +
\frac18 U_i U_k F_{mn}(W+P) \right\}\,.
\label{DE05}
\end{equation}
This contribution is linear in the state functions of the Dark Energy, $P\equiv P^{({\rm DE})}$ and $W\equiv W^{({\rm DE})}$, and is quadratic in the Maxwell tensor $F_{ik}$.

\subsubsection{An Illustration: Dark Epochs in the Universe History Caused by  Striction-Type coupling}

In \cite{BD2014} we considered three illustrations of the formula (\ref{22varco}) in the framework of the model with Archimedean type force acting on the DM particles in the DE reservoir (see \cite{Arc1,Arc2,Arc3} for details): first, for the de Sitter-type solution with the cosmological constant $\Lambda$; second, for an anti-Gaussian solution, which describes a specific bounce in the Universe evolution; third, for a super-exponential expansion of the Universe. For the first case the DE pressure is constant, $P=-\Lambda$, thus the refraction index, phase and group velocities are constant; the Dark Epochs are absent.
When the DE pressure depends on time, it is more convenient to rewrite (\ref{22varco}) as
 \begin{equation}
n^2(t) = \frac{\alpha}{\beta}  \frac{\left(P-P_{(\varepsilon)}\right)}{\left(P - P_{(\mu)}\right)} \,,  \quad P_{(\varepsilon)} \equiv \frac{\varepsilon_{(0)}}{\alpha} \,, \quad
P_{(\mu)} \equiv \frac{1}{\mu_{(0)}\beta} \,.
\label{23varco}
\end{equation}
For an illustration, let both $\alpha$ and $\beta$ be positive, and $P_{(\varepsilon)}> P_{(\mu)}$. Then
the function $n^2(t)$ is negative when $P_{(\mu)}<P<P_{(\varepsilon)}$. Again we deal with Dark Epochs, when the electromagnetic waves can not propagate, but now the striction-type interaction with the Dark Energy is the origin of this phenomenon. The function $n^2(t)$ can change the sign at the moment $t=t_{*}$, for which $n^2(t_{*})=0$, or at the moment $t=t_{**}$, when $n^2(t_{**})=\infty$.
In the first case the DE pressure coincides with its critical value $P_{(\varepsilon)}$, and $V_{({\rm ph})}(t_{*})= \infty$, $V_{({\rm gr})}(t_{*})=0$.
In the second case the DE pressure coincides with its critical value $P_{(\mu)}$, and $V_{({\rm ph})}(t_{**})= V_{({\rm gr})}(t_{**})=0$.
Clearly, the refraction index is constant, if $P_{(\varepsilon)}{=}P_{(\mu)}$, i.e., when $n^2_{(0)} = \frac{\alpha}{\beta}$.

\section{Model 6: Piezo-Type Coupling via a Scalar Dark Energy}

The piezo-electric and piezo-magnetic effects are well-known in classical electrodynamics of anisotropic materials, in which the electric and magnetic fields, respectively, can appear under the influence of pressure (strain) \cite{ED2}. In the covariant electrodynamics of continua these effects can be described by the Lagrangian
\begin{equation}
L_{(6)} = \frac12 {\cal D}^{ikpq} F_{ik} \ {\cal P}_{pq} \,,
\label{piezo1}
\end{equation}
into which the Maxwell tensor and the pressure tensor enter linearly.
The piezo-electric and piezo-magnetic coefficients are encoded
in the tensor ${\cal D}^{ikpq}$.
The corresponding contributions into the induction tensor
\begin{equation}
{\cal H}^{ik}_{(6)} = {\cal D}^{ikpq}{\cal P}_{pq}
\label{2induc11}
\end{equation}
does not contain $F_{ik}$ and has to be included into the total tensor of spontaneous polarization - magnetization.
The tensor ${\cal D}^{ikpq}$ is symmetric with respect to the indices in the last pair $pq$, and is skew-symmetric with respect to $ik$.
Since the symmetric pressure tensor ${\cal P}_{pq}$ is orthogonal to the velocity
four-vector $U^i$, we deal with the coefficients, which satisfy the relationships
\begin{equation}
{\cal D}^{ikpq} U_p = 0 = {\cal D}^{ikpq} U_q \,.
 \label{d2}
\end{equation}
Using the unit velocity four-vector $U^i$, we can decompose the tensor ${\cal D}^{ikpq}$ as
\begin{equation}
{\cal D}^{ikpq} = d^{i(pq)} U^k - d^{k(pq)} U^i - \epsilon^{ik}_{\ \ ls}U^s h^{l(pq)}  \,,
 \label{d3}
\end{equation}
where the piezo-electric coefficients $d^{i(pq)}$ and piezo-magnetic coefficients $h^{l(pq)}$ are introduced by
\begin{equation}
d^{i(pq)} \equiv {\cal D}^{ikpq} U_k \,, \quad h^{l(pq)} \equiv \frac12 \epsilon^{ls}_{\ \ ik} {\cal D}^{ikpq} U_s \,.
 \label{d4}
\end{equation}
Both these tensors are symmetric with respect to $pq$, and  are orthogonal to $U^i$
\begin{equation}
d^{i(pq)} U_i {=} 0 {=}  d^{i(pq)} U_p  \,, \quad h^{l(pq)} U_l {=}  0 {=} h^{l(pq)} U_p \,.
 \label{d5}
\end{equation}
In general case, the DE influence can be
characterized by 18 piezo-electric coefficients $d^{i(pq)}$ and/or
by 18 piezo-magnetic coefficients $h^{l(pq)}$. When the Dark
Energy is spatially isotropic, ${\cal D}^{ikpq}=0$, i.e., only in the framework of anisotropic cosmological models the DE-induced piezo-effect can be activated.
And finally, the tensor
\begin{equation}
T^{(6)}_{ik} = 2{\cal D}^{mnpq}F_{mn} \left[ \frac14 g_{ik}  {\cal P}_{pq}
{+}   {\cal B}_{ikls}\Delta^l_p \Delta^s_q \right]
{-}
F_{mn} {\cal P}_{ls} \frac{\delta}{\delta g^{ik}}\left({\cal D}^{mnpq} \Delta^l_p \Delta^s_q \right)
 \label{S99}
\end{equation}
is the piezo-contribution into the total stress-energy tensor (see \cite{BD2014} for details of calculations). It is interesting to apply this formalism to the model of DE-induced piezo-magnetic effects in the Bianchi-I anisotropic Universe with magnetic field; this work is in progress.

\section{Model 7: Pyro-Type Coupling via a scalar Dark Energy}

Classical pyro-electricity is the effect of  polarization of a material under the influence of temperature variation; similarly, one deals with effect of pyro-magnetism, if the magnetization appears in the material with varying temperature. If the internal energy density ${\cal E}$ of a material is a function of the temperature only, ${\cal E}(T)$, one can link the rates $\dot{{\cal E}}$ and
$\dot{T}$ by the relation $\dot{{\cal E}}(T)= \frac{d{\cal E}}{dT} \dot{T}$. In other words, one can say that pyro-effects are the results of variation of the internal energy of pyro-active materials.

In cosmology, the energy density of the Dark Energy (treated as a fluid) also depends on time, and we could speak about DE - induced pyro-electric and pyro-magnetic phenomena. Mathematically, these phenomena can be described by the Lagrangian
\begin{equation}
L_{(7)} = \frac12  \pi^{ik} F_{ik}  DW \,,
\label{pyro1}
\end{equation}
linear in the Maxwell tensor and linear in the convective derivative of the DE energy density $W \equiv W^{({\rm DE})}$.
Similarly to the case with piezo-phenomena, we obtain that there is a contribution
\begin{equation}
{\cal H}^{ik}_{(7)} \equiv \pi^{ik} DW
\label{pyro2}
\end{equation}
into the total tensor of spontaneous polarization-magnetization.
The skew-symmetric tensor $\pi^{ik}$ can be represented as
\begin{equation}
\pi^{ik}= \pi^i U^k - \pi^k U^i - \epsilon^{ik}_{\ \ mn} \mu^{m}
U^n \,,
 \label{pi03}
\end{equation}
thus introducing the pyro-electric  $\pi^i$ and pyro-magnetic
$\mu^m$ coefficients, which are orthogonal to $U^i$. In general case there are three pyro-electric and three
pyro-magnetic coefficients; in a spatially isotropic medium all the pyro-coefficients vanish.
The pyro-contribution into the total stress-energy tensor can be written as
$$
T^{(7)}_{ik} =
 \left[\frac12 g_{ik} F_{mn}\pi^{mn}
{-}F_{mn} \frac{\delta}{\delta g^{ik}} \pi^{mn} \right] DW  +
$$
\begin{equation}
 + \left[W\left(\frac12 g_{ik}+U_iU_k \right) - 2 {\cal B}_{ikls} U^lU^s \right] \nabla_j \left[U^j F_{mn}\pi^{mn}\right]
- \frac12  F_{mn}\pi^{mn} U_{(i} \nabla_{k)} W \,, \label{pi195}
\end{equation}
(see \cite{BD2014} for details of calculations). This contribution is non-vanishing for anisotropic cosmological models, e.g., for the Bianchi-I model with magnetic field (the work on the corresponding application also is in progress).

\section{Model 8: Dynamo-Optical Interactions Associated with  Dark Energy}

\subsection{Irreducible Representation of Basic Quantities}

The Lagrangian of this model does not contain the pseudoscalar (axion) field, and thus dynamo-optical interactions of this type can be attributed to the coupling of electromagnetic field
to the Dark Energy. As usual, we consider the Lagrangian, which includes terms linear and quadratic in the Maxwell tensor.
\begin{equation}
L_{(8)}=
\left[\frac12 A^{mnpq}  F_{pq} +  \frac{1}{4} X^{mnikpq} F_{ik}F_{pq}\right] \nabla_m U_n \,.
\label{8M1}
\end{equation}
The contribution to the induction tensor is now of the form
\begin{equation}
H^{ik}_{(8)}=
A^{mnik} \nabla_m U_n +   X^{mnikpq} (\nabla_m U_n) F_{pq}   \,.
\label{8M2}
\end{equation}
Clearly, the first term in (\ref{8M2}) contributes into the spontaneous polarization-magnetization tensor
${\cal H}^{ik}_{(8)}= A^{mnik} \nabla_m U_n$; the second terms gives the susceptibility tensor $\chi^{ikpq}_{(8)} = X^{mnikpq} \nabla_m U_n $.

As it was shown in \cite{BL2014}, generally, the tensor $A^{mnik}$ contains only two independent coupling constants
$$
A^{mnik} = \pi_{(8)}  g^{iknl} U^m U_l - \mu_{(8)} \Delta^{ikmn} \,,
$$
\begin{equation}
g^{iknl} \equiv g^{in}g^{kl}-g^{il}g^{kn} \,, \quad \Delta^{ikmn} \equiv \Delta^{im}\Delta^{kn}-\Delta^{in}\Delta^{km} \,.
\label{M5}
\end{equation}
With this tensor $A^{mnik}$, the spontaneous polarization four-vector
\begin{equation}
{\cal P}^{i}_{(8)} \equiv {\cal H}^{ik}_{(8)}U_k =  \pi_{(8)} DU^i
\label{8M5}
\end{equation}
is proportional to the acceleration four-vector $DU^i$, the only true four-vector, which can be formed using $\nabla_iU_k$, $\Delta^{ik}$, $\eta^{ikm}$ and $U^i$.
Searching for ${\cal M}^{i}$ we can find only one natural pseudo four-vector, $\omega^{i}{=} -2\omega^{*ik}U_k$; one can check directly that with given $A^{mnik}$
the spontaneous magnetization four-vector
\begin{equation}
{\cal M}^{i}_{(8)} \equiv {\cal H}^{*ik}_{(8)}U_k =
\mu_{(8)} \epsilon^{ikpq} U_k \omega_{pq}
\label{8M6}
\end{equation}
is proportional to $\omega^{i}$, the angular velocity of the medium rotation.

The tensor $X^{lsikmn}$ is reconstructed in \cite{BL2014} as follows:
$$
X^{lsikmn} =
\frac12 \left(\alpha_{(81)} {-} \frac13 \alpha_{(86)} \right)
\Delta^{ls}\left(g^{ikmn}-\Delta^{ikmn} \right) +
\frac14 \alpha_{(86)} U_p U_q \left[g^{iklp}g^{mnsq} +
g^{mnlp}g^{iksq} \right]+
$$
\begin{equation}
+ \frac12 \left(\gamma_{(81)} {-} \frac13 \gamma_{(86)} \right)
\Delta^{ls}\Delta^{ikmn} - \frac12 \gamma_{(86)} \
\eta^{ik(l} \eta^{s)mn} - \nu_{(8)} U^l \left\{\Delta^{iks[m}U^{n]}
+  \Delta^{mns[i}U^{k]}\right\}
\,.
\label{X}
\end{equation}
The corresponding contributions into the dielectric permittivity, magnetic impermeability and magneto-electric tensor read
\begin{equation}
\varepsilon^{ik}_{(8)} = \Delta^{ik} \alpha_{(81)} \Theta
 + \alpha_{(86)} \sigma^{ik} \,, \quad
{\left(\mu^{-1}\right)}^{ik}_{(8)} = \Delta^{ik} \gamma_{(81)} \Theta {+} \gamma_{(86)} \sigma^{ik} \,, \quad
\nu^{pm}_{(8)} = \nu_{(8)} \eta^{pml} DU_l \,.
\label{8nu}
\end{equation}
Let us emphasize that the vorticity tensor does not appear in $X^{lsikmn}$ due to the symmetry of this tensor.

\subsection{An Illustration: Dynamo-Optical Interactions with Dark Energy Provoked by Gravitational pp-Waves}

As an illustration, let us consider again the pp-wave symmetric space-time with the metric (\ref{GWmetric}) (see Subsection \ref{GW1}).
Let us assume, that before the gravitational wave incoming ($u<0$) the spatially isotropic electrodynamic medium was characterized by permittivity parameters $\varepsilon$ and $\mu$, the electric field was absent, and the constant  magnetic field was orthogonal to the direction of the gravitational pp-wave propagation. For this metric and for the velocity four-vector $U^i=\delta^i_0$, the acceleration four-vector and the vorticity tensor are equal to zero, $DU^i=0$, $\omega_{ik}=0$.
The expansion scalar is equal to
\begin{equation}
\Theta = \frac{\sqrt{2}\,L^{\prime}(u)}{L} \,,
\label{GW0011}
\end{equation}
and the shear tensor can be written as
\begin{equation}
\sigma^k_i  = \frac{\Theta}{2}  \left(\frac13
\Delta_i^k - \delta_i^1 \delta^k_1 \right)
+ \frac{\beta^{\prime}}{\sqrt2} \left(\delta_i^2 \delta^k_2
{-} \delta_i^3 \delta^k_3 \right)\,.
\label{GW002}
\end{equation}
Since the acceleration and vorticity are absent for such a velocity field, the coupling constants $\pi_{(8)}$, $\mu_{(8)}$ and $\nu_{(8)}$ become the hidden parameters of the model
(see (\ref{8M5}), (\ref{8M6}) and (\ref{8nu})).
As it was shown in \cite{AB2016}, for this configuration the longitudinal (with respect to the propagation direction of the gravitational waves) components of the electric and magnetic fields remain vanishing. The transversal components happen to be deformed.

\subsubsection{Exact Solutions for the Transversal Electric and Magnetic Fields}

Transversal magnetic field dynamo-optically interacting with DE in the field of gravitational pp-wave generates the transversal electric field:
\begin{equation}
E^2(u)= \frac{B_3(0)}{\Delta_{(+)}}\left\{\frac{1}{\mu}\left(1{-}e^{{-}2\beta} \right){-} e^{{-}2\beta}\left[ \frac{\sqrt2 L^{\prime}}{L}\left(\gamma_{(81)}{+}\frac16 \gamma_{(86)} \right)
 -\frac{\beta^{\prime}}{\sqrt2} \gamma_{(86)} \right] \right\}    \,,
\label{S23}
\end{equation}
\begin{equation}
E^3(u)= -\frac{B_2(0)}{\Delta_{(-)}}\left\{\frac{1}{\mu}\left(1-e^{2\beta} \right)-e^{2\beta}\left[  \frac{\sqrt2 L^{\prime}}{L} \left(\gamma_{(81)}+\frac16 \gamma_{(86)} \right)
 + \frac{\beta^{\prime}}{\sqrt2} \gamma_{(86)} \right] \right\}   \,.
\label{S24}
\end{equation}
Clearly, the electric field components vanish at the initial moment, $E^2(0)=0$ and $E^3(0)=0$.
The magnetic field is deformed:
\begin{equation}
B_2(u)= \frac{L^2B_2(0)}{\Delta_{(-)}}\left\{\left(\varepsilon e^{2\beta} -\frac{1}{\mu}\right) + e^{2\beta}\left[ \frac{\sqrt2 L^{\prime}}{L} \left(\alpha_{(81)}+\frac16 \alpha_{(86)} \right) -
\frac{\beta^{\prime}}{\sqrt2} \alpha_{(86)} \right] \right\}  \,,
\label{11S3}
\end{equation}
\begin{equation}
B_3(u)= \frac{L^2B_3(0)}{\Delta_{(+)}}\left\{\left(\varepsilon e^{-2\beta} -\frac{1}{\mu}\right) + e^{-2\beta}\left[  \frac{\sqrt2 L^{\prime}}{L} \left(\alpha_{(81)}+\frac16 \alpha_{(86)} \right) +
\frac{\beta^{\prime}}{\sqrt2} \alpha_{(86)} \right] \right\} \,.
\label{12S3}
\end{equation}
The denominators in the formulas (\ref{S23})-(\ref{12S3})
\begin{equation}
\Delta_{(+)}(u) \equiv L^2\left\{\left(\varepsilon {-} \frac{1}{\mu} \right) + \frac{1}{\sqrt2} \beta^{\prime}(\alpha_{(86)}{+} \gamma_{(86)})
+   \frac{\sqrt2 L^{\prime}}{L} \left[(\alpha_{(81)}{-}\gamma_{(81)})+\frac16(\alpha_{(86)}{-}\gamma_{(86)}) \right] \right\} \,,
\label{S9a}
\end{equation}
\begin{equation}
\Delta_{(-)}(u) \equiv L^2\left\{\left(\varepsilon {-} \frac{1}{\mu} \right) - \frac{1}{\sqrt2} \beta^{\prime}(\alpha_{(86)}{+} \gamma_{(86)})
+  \frac{\sqrt2 L^{\prime}}{L} \left[(\alpha_{(81)}{-}\gamma_{(81)})+ \frac16(\alpha_{(86)}{-}\gamma_{(86)}) \right] \right\}
\label{S11a}
\end{equation}
can take zero values at some moments of the retarded time (say, $u^*$), thus providing the anomalies
in the responses of the electromagnetic field.

\subsubsection{Explicit Example of Anomaly}

For the illustration of anomaly, we consider the function $\Delta_{(+)}(u)$
for the Petrov metric with
\begin{equation}
L^2 = \cos{ku} \cdot \cosh{ku} \,, \quad 2\beta = \log{\left[\frac{\cos{ku}}{\cosh{ku}}\right]} \,,
\label{b59}
\end{equation}
(see \cite{Petrov}). Since now we deal with the explicit functions
\begin{equation}
\Theta(u) = \frac{\sqrt2 L^{\prime}}{L} = \frac{k}{\sqrt2}\left(\tanh{ku} - \tan{ku} \right) \,, \quad \Theta(0) = 0 \,, \label{q59}
\end{equation}
\begin{equation}
\beta^{\prime}(u) = - \frac{k}{2} \left(\tanh{ku} + \tan{ku} \right) \,, \quad \beta^{\prime}(0)=0 \,,
\label{q57}
\end{equation}
one can state the following: first, $\Delta_{(+)}(0)=\left(\varepsilon {-} \frac{1}{\mu} \right) >0$; second,
\begin{equation}
\Delta_{(+)}\left(\frac{\pi}{2k}\right) = - \cosh{\frac{\pi}{2}}\left[ (\alpha_{(81)}{-}\gamma_{(81)})+ \frac13(2\alpha_{(86)}{+}\gamma_{(86)}) \right]
 \,.
\label{S31}
\end{equation}
When $ku=\frac{\pi}{2}$ we obtain that $L=0$, i.e., the metric degenerates, and we have to deal with the admissible interval $0\leq u < \frac{\pi}{2k}$.
Clearly, if the coupling parameters are linked by the inequality $\frac13(2\alpha_{(86)}{+}\gamma_{(86)}) > \gamma_{(81)} {-} \alpha_{(81)}$, we obtain that $\Delta_{(+)}\left(\frac{\pi}{2k} \right)<0$.
This means that the function $\Delta_{(+)}(u)$, which starts with positive value at $u=0$ and finishes with negative value at $u=\frac{\pi}{2k}$, takes zero value inside the admissible interval at the moment $u=u^{**}$. The  component $E^2$ of the electric field is infinite at this moment, while the component $E^3$ remains finite.

\section{Model 9: Non-minimal Coupling of Photons to the Dark Fluid}

\subsection{Mathematical Aspects of the Model}

\subsubsection{The Lagrangian}

One speaks about non-minimal coupling between some fields, when the interaction Lagrangian $\pounds_{({\rm interaction})}$ contains the Riemann tensor $R^i_{\ kmn}$ and/or its convolutions, Ricci tensor $R_{ik}$ and Ricci scalar $R$. The story of elaboration of the theory of non-minimal coupling deserves a special review; here we mention only the theory on non-minimal coupling of the electromagnetic field to gravity, and do it only in the context of photon interactions with the Dark Fluid.
For this particular task we consider the following contribution into the Lagrangian:
\begin{equation}
L_{(9)} = \frac{1}{4}
{\cal R}^{ikmn} F_{ik}F_{mn} {+}  \frac{1}{4} {\chi}^{ikmn}_{({\rm
Axion})}  \phi \ F_{ik} F^{*}_{mn} {-} \frac{1}{2} \eta_{(1)}
F_{ik} \nabla^i \phi \  R^{kn} \nabla_n \phi {-} \frac12 \eta_{({\rm A})} R  \phi^2 \,.
\label{a9}
\end{equation}
The tensor  ${\cal R}^{ikmn}$
\begin{equation}
{\cal R}^{ikmn} =  q_1 R g^{ikmn} + q_2 \Re^{ikmn} + q_3 R^{ikmn}
\,, \label{sus1}
\end{equation}
contains three coupling parameters $q_1$, $q_2$ and $q_3$ in front of tensors
\begin{equation}
g^{ikmn} \equiv \frac{1}{2}(g^{im}g^{kn} {-} g^{in}g^{km}) \,,
\label{rrr}
\end{equation}
\begin{equation}
\Re^{ikmn} \equiv \frac{1}{2} (R^{im}g^{kn} {-} R^{in}g^{km} {+}
R^{kn}g^{im} {-} R^{km}g^{in}) \,, \label{rrrr}
\end{equation}
and $R^{ikmn}$, respectively (in this Section we put the multiplier $\frac12$ in definition of $g^{ikmn}$, keeping in mind historical motives). The quantity ${\cal R}^{ikmn}$  can be indicated as a non-minimal three-parameter susceptibility tensor \cite{NM7}.
The quantity ${\chi}^{ikmn}_{({\rm Axion})}$ is given by
\begin{equation}
{\chi}^{ikmn}_{({\rm Axion})} {=} Q_1 R g^{ikmn} {+} Q_2\Re^{ikmn}
{+} Q_3 R^{ikmn} \,, \label{sus2}
\end{equation}
with coupling constants $Q_1$, $Q_2$ and $Q_3$. This part of the Lagrangian describes the non-minimal
interaction of the electromagnetic field with gravitation, mediated by the coupling to the axionic Dark Matter. The term with the coupling constant $\eta_{(1)}$ in front, is linear in the Maxwell tensor; the last term does not contain $F_{ik}$.
Standardly, the tensors ${\cal R}^{ikmn}$ and  ${\chi}^{ikmn}_{({\rm Axion})}$,
are skew-symmetric with respect to transpositions in the pairs of the indices $ik$ and $mn$. In addition, one can see that ${\cal R}^{ikmn}= {\cal R}^{mnik}$. As for the tensor ${\chi}^{ikmn}_{({\rm
Axion})}$, we require that it is symmetric with respect to dualization procedure
\begin{equation}
{}^{*}{\chi}^{ikmn}_{({\rm Axion})} = {\chi}^{*ikmn}_{({\rm Axion})}   \ \Leftarrow \cdot \Rightarrow  \ {}^{*}{\chi}^{*ikmn}_{({\rm Axion})} = - {\chi}^{ikmn}_{({\rm Axion})}  \,.
\label{sus27}
\end{equation}
Since
\begin{gather}
{}^{*}{g}^{*ikmn} = - g^{ikmn}  \,,\label{sus29}\\
{}^{*}{\Re}^{*ikmn} =  {\Re}^{ikmn} - R g^{ikmn}  \,,
\label{sus30}\\
{}^{*}{R}^{*ikmn} = -R^{ikmn}+ 2 \Re^{ikmn}- Rg^{ikmn}  \,,
\label{sus31}
\end{gather}
one can conclude that the symmetry condition (\ref{sus27}) leads to the restriction  $Q_2{+}Q_3=0$.

\subsubsection{Contributions into the Master Equations of Electromagnetic Field }

For the model with the interaction Lagrangian (\ref{a9}), the contribution into the induction tensor is
\begin{equation}
H^{ik}_{(9)} =  {-} \eta_{(1)}
\nabla^{[i} \phi \  R^{k]n} \nabla_n \phi  + {\cal R}^{ikmn} F_{mn} +  {\chi}^{ikmn}_{({\rm
Axion})}  \phi \ F^{*}_{mn} \,. \label{eld2}
\end{equation}
The first term forms the contribution into the spontaneous polarization-magnetization tensor
\begin{equation}
{\cal H}^{ik}_{(9)} =  {-} \eta_{(1)}
\nabla^{[i} \phi \  R^{k]n} \nabla_n \phi \,. \label{eld222}
\end{equation}
The second and third terms relate to the non-minimal susceptibility tensor
\begin{equation}
\chi^{ikmn}_{(9)} =  {\cal R}^{ikmn} +  \phi {\chi}^{*ikmn}_{({\rm
Axion})}  \,, \label{eld3}
\end{equation}
which gives the following permittivity and cross-effect tensors:
\begin{equation}
\varepsilon^{im}_{(9)} = \Delta^{im} + 2 \left[{\cal R}^{ikmn} +  \phi {\chi}^{*ikmn}_{({\rm
Axion})} \right]U_k U_n \,,
\label{Re}
\end{equation}
\begin{equation}
{(\mu^{-1})_{pq}}_{(9)} = \Delta_{pq} - \frac{1}{2} \eta_{pik}\left[{\cal R}^{ikmn} +  \phi {\chi}^{*ikmn}_{({\rm
Axion})} \right] \eta_{mnq} \,, \label{Rmu}
\end{equation}
\begin{equation}
{\nu_{p}^{\ m}}_{(9)} =  \eta_{pik} \left[{\cal R}^{ikmn} +  \phi {\chi}^{*ikmn}_{({\rm
Axion})} \right] U_n \,.
\label{RnuR}
\end{equation}
These tensors depend on five effective non-minimal coupling parameters $q_1$, $q_2$, $q_3$, $Q_1$, $Q_3$.

\subsubsection{Contributions into the Master Equation for the Pseudoscalar Field }

The non-minimally modified equation for the pseudoscalar field
\begin{gather}
\nabla_m \left[ \left( g^{mn} {+} \Re^{mn}_{({\rm A})} \right)
\nabla_n \phi \right] {+} \left[V^{\prime}(\phi^2){+} \eta_{({\rm A})} R \right] \phi = - \frac{1}{4\Psi^2_0} F^{*}_{mn}\left(F^{mn}{+} {\chi}^{ikmn}_{({\rm A})} F_{ik} \right) \,,
\label{eqaxi13}
\end{gather}
contains three new elements. First,
we obtain the additional term
\begin{equation}
{\cal J}_{(9)} = -\frac{1}{4\Psi^2_0} {\chi}^{ikmn}_{({\rm
Axion})} F_{ik} F^{*}_{mn}
\label{Rnu}
\end{equation}
in the right-hand side of this equation. Second, the tensor $\tilde{g}^{mn}= g^{mn} {+} \Re^{mn}_{({\rm A})}$ with
\begin{equation}
\Re^{mn}_{({\rm A})} \equiv \frac{1}{2} \eta_{(1)}
\left(F^{ml}R^{n}_{\ l} + F^{nl}R^{m}_{\ l} \right) \label{sus3}
\end{equation}
plays the role of effective metric for pseudoscalar waves (see \cite{BZD1,BZD2,BZD3} for details). Third, the term $ \eta_{({\rm A})} R$ describes the curvature induced contribution to the square of effective mass of the pseudoscalar field.

\subsubsection{Non-minimal Extension of the Gravity Field Equations}

Non-minimal contributions into the right-hand side of the gravity field equations are very sophisticated; we can represent them in the following form:
\begin{equation}
T^{(9)}_{ik} = q_1 T^{(91)}_{ik} + q_2 T^{(92)}_{ik}+ q_3
T^{(93)}_{ik} + \left(Q_1-\frac{1}{2}Q_3
\right) T^{(94)}_{ik} + Q_3 T^{(95)}_{ik} +
\eta_{(1)} T^{(96)}_{ik} + \eta_{({\rm A})} T^{(97)}_{ik} \,, \label{9decompEM}
\end{equation}
where the listed contributions to the stress-energy tensor read
\begin{gather}
T^{(91)}_{ik} = \frac{1}{2} \left[ \nabla_{i} \nabla_{k} - g_{ik}
\nabla^l \nabla_l \right] \left[F_{mn}F^{mn} \right]
- R F_{im}F_{k}^{ \ m}  - \frac{1}{2} F_{mn}F^{mn}
\left(R_{ik}{-}\frac{1}{2}Rg_{ik}\right)
 \,, \label{T1}
\end{gather}
\begin{gather}
T^{(92)}_{ik} = -\frac{1}{2}g_{ik} \left[\nabla_{m}
\nabla_{l}\left(F^{mn}F^{l}_{\ n} \right) - R_{lm} F^{mn} F^{l}_{\
n} \right]- F^{ln} \left(R_{il}F_{kn} + R_{kl}F_{in} \right)- \frac{1}{2}
\nabla^m \nabla_m \left(F_{in} F_{k}^{ \ n}\right)+{}\nonumber\\
{}+\frac{1}{2}\nabla_l \left[ \nabla_i \left( F_{kn}F^{ln} \right)
{+} \nabla_k \left(F_{in}F^{ln} \right) \right] {-} R^{mn} F_{im}
F_{kn} \,, \label{T2}
\end{gather}
\begin{gather}
T^{(93)}_{ik} {=} \frac{1}{4}g_{ik} R^{mnls}F_{mn}F_{ls}{-}
\frac{3}{4} F^{ls} \left(F_{i}^{\ n} R_{knls} {+} F_{k}^{\
n}R_{inls} \right) - \frac{1}{2} \nabla_{m}
\nabla_{n} \left[ F_{i}^{ \ n}F_{k}^{ \ m} + F_{k}^{ \ n} F_{i}^{
\ m} \right] \,, \label{T3}
\end{gather}
\begin{gather}
T^{(94)}_{ik}  \equiv  \frac{1}{2} \left[ \nabla_{i}
\nabla_{k} {-} g_{ik} \nabla^l \nabla_l \right] \left[\phi
\Fst_{mn}F^{mn} \right] -\frac{1}{2} R_{ik}\phi \Fst_{mn}F^{mn} \,, \label{calT1}
\end{gather}
\begin{gather}
T^{(95)}_{ik}  \equiv {-}\frac{1}{2} \nabla_{m} \nabla_{n}
\left[ \phi \left(\Fst_i^{\ n} {F_k}^{m} {+} \Fst_k^{\ n}
{F_i}^{m} \right)\right]+ \frac{1}{4} \phi \Fst^{mn} \left(F_{il} R^{l}_{ \ kmn} {+}
F_{kl} R^{l}_{\ imn} \right) \,, \label{calT3}
\end{gather}
\begin{gather}
T^{(96)}_{ik} \equiv  \frac{1}{2}g_{ik} \left(R^l_n {-}
\nabla^l \nabla_n \right) \left( F^{nm} \nabla_m \phi \nabla_l
\phi \right) +\frac{1}{2} R^l_n \nabla_l \phi \left(F_i^{\ n} \nabla_k \phi
{+} F_k^{\ n} \nabla_i \phi \right) +{}\nonumber \\
{}+\frac{1}{4}\nabla^l \nabla_l \left[ \nabla_m \phi \left(
 F^m_{\ \ i} \nabla_k \phi {+}  F^m_{\ \ k} \nabla_i \phi
\right)\right]+ \frac{1}{4}\nabla^l \left[ \nabla_i \left(F_k^{\ m}\nabla_m
\phi \nabla_l \phi \right) {+} \nabla_k \left(F_i^{\ m}\nabla_m
\phi \nabla_l \phi \right) \right] +{}\nonumber \\
{}+\frac{1}{4}\nabla_m \left[ \nabla_i \left( F^{mn}\nabla_k \phi
\nabla_n \phi  \right) {+} \nabla_k \left( F^{mn}\nabla_i \phi
\nabla_n \phi  \right) \right] +{}\nonumber \\
{}+\frac{1}{2} F^{mn}   \left(R_{in}\nabla_k \phi {+}
R_{kn}\nabla_i \phi \right)\nabla_m \phi + \frac{1}{2}\left(R^m_i F^n_{\ k} {+} R^m_k F^n_{\ i} \right)
\nabla_m \phi \nabla_n \phi \,, \label{calT4}
\end{gather}
\begin{equation}\label{calT7}
T^{(97)}_{ik}= \left(\nabla_i \nabla_k - g_{ik} \nabla_m
\nabla^m \right){\phi}^2 - \left(R_{ik}{-}\frac{1}{2}Rg_{ik}\right) \phi^2\,.
\end{equation}
Only the last term does not contain the Maxwell tensor.

\subsection{Three Illustrations of the Non-minimal Model}

\subsubsection{Cosmological Dark Epochs Produced by Interacting DM and DE}

In the framework of non-minimal Einstein-Maxwell theory applied to the isotropic cosmological FLRW model,
the square of an effective refraction index can be found as follows (see \cite{BBL2012}):
\begin{equation}
n^2(t) = \frac{1 - 2 (3q_1 {+} 2q_2 {+} q_3)\frac{\ddot{a}}{a} -2
(3q_1 {+} q_2) \left(\frac{\dot{a}}{a}\right)^2}{1 - 2(3q_1 {+}
q_2)\frac{\ddot{a}}{a} -2 (3q_1 {+} 2q_2{+} q_3)
\left(\frac{\dot{a}}{a}\right)^2 }\,.
\label{R19}
\end{equation}
For the illustration, we study the model, for which ${\cal R}^{ikmn}g_{im}g_{kn} \equiv {\cal R}{=}0$; it is possible when
$6q_1{+}3q_2{+}q_3{=}0$.
Then we consider a scale factor $a(t)$ presented  by a
stretched exponential function
\begin{equation}
a(t)=a_0 \exp\{ ( \Gamma t )^{\nu}\}\,,
\label{Kohl1}
\end{equation}
where $\Gamma$ and $\nu$ are some constants.
Such a  function was introduced by Kohlrausch \cite{Kohl} in 1854.  If
$\nu {=}1$ the function (\ref{Kohl1}) coincides with the standard de
Sitter exponent, if we put $H_0=\Gamma$. If $\nu{=}2$ we deal with an anti-Gaussian
function studied in \cite{Arc1} as an exact solution of
a model with Archimedean-type interaction between DE and DM. The stretched exponent was used in \cite{O1}
in the context of generalized Chaplygin
gas models.
When the scale factor is described by (\ref{Kohl1}), and $6q_1{+}3q_2{+}q_3=0$, we obtain immediately that
\begin{equation}
n^2(t)= \frac{(\Gamma t )^{2-\nu}+ \nu (\nu{-}1)\Gamma^2 \bar{Q}}{(\Gamma t
)^{2-\nu}- \nu
(\nu{-}1)\Gamma^2 \bar{Q}} \,, \quad  \bar{Q} \equiv 2 (3q_1 {+} q_2) \,.  \label{Kohl3}
\end{equation}
Clearly, for the de Sitter law, $\nu{=}1$, the refraction index is equal to one $n^2(t) \equiv
1$; there are no Dark Epochs. For the late-time Universe evolution the refraction index in vacuum has to take (asymptotically) the value $n{=}1$. One can see from (\ref{Kohl3}) that the function $n^2(t)$ tends to one asymptotically at $t \to \infty$, only when $\nu<2$.
For $1<\nu<2$, either the numerator (for $\bar{Q}<0$), or denominator (for $\bar{Q}>0$) can take zero values. For instance, if $\bar{Q}$ is positive,
$n^2<0$ during the interval $0<t<t_{*}$, where $t_{*}{=}\frac{1}{\Gamma}[\nu(\nu{-}1)\Gamma^2 \bar{Q}]^{\frac{1}{2{-}\nu}}$. At $t{=}t_{*}$ the refraction index
is infinite, the phase and group velocities are equal to zero; the moment $t{=}t_{*}$ is the finishing point of a Dark Epoch started at $t{=}0$.

\subsubsection{Example of Regular Static Solution with Spherical Symmetry}

Non-minimal coupling of photons to the gravity field is shown to form regular and quasi-regular field configurations (see, e.g.,  \cite{Reg1,Reg2,Reg3,Reg4,Reg5}).
When we deal with an electrically charged (with the total charge $Q$) spherically symmetric object (star, monopole, black hole, etc.) with the metric (\ref{metric2}), it is useful to operate with the dimensionless radial variable $x{=}\frac{r}{r_{Q}}$, where
 $r_{Q}\equiv \sqrt{G}|Q|$. In \cite{Reg2} we have found the exact explicit solution to the master equations of non-minimal model for the case, when $2|q_1|{=}r^2_{Q}$; the corresponding solution for the radial electric field is of the form:
\begin{equation}
E(x) = \frac{Q}{2r^2_{Q}(1+x^2)} \left[ 1-x^2 + \sqrt{x^4
+ 2x^2 +5} \right] \,. \label{z3}
\end{equation}
This solution has the standard Coulombian asymptote $E(r) \to \frac{Q}{r^2}$ at $r \to \infty$. In the center, at $r=0$,
$E(0){=}\frac{Q}{r^2_{Q}} \frac{(\sqrt{5}{+}1)}{2}$, i.e., we deal with the solution regular at the center, the value $E(0)$ being proportional to the so-called  ``golden section'' $\phi \equiv \frac{\sqrt5 {+}1}{2}$. The metric function $\sigma(x)$ also is found explicitly
\begin{equation}
\sigma(x) =  \exp\left\{- \ \frac{ 3  + (1-x^2) \sqrt{x^4
+ 2x^2 +5}+ x^4}{2 (1+x^2)^2} \right\} \,.
\label{si}
\end{equation}
Asymptotic value of this function is $\sigma(\infty)=1$; the value at the center is finite $\sigma(0)=\exp\{-(1+\phi) \}$.
The second metric function, $N(x)$, is found in quadratures
\begin{equation}
N(x) = \frac{1}{2x \sigma(x)} \int_0^{x}
d \xi \ \sigma(\xi) \left[  \xi^2 + 3
- \sqrt{\xi^4 +2 \xi^2 +5} \right] \,.
\label{Nsi}
\end{equation}
Clearly,  $N(\infty) =1$ and $N(0) {=} \frac{3-\sqrt5}{2}$, so that $1-N(0) {=} \frac{1}{\phi} \equiv \phi-1$.
The solutions for the metric functions $\sigma(r)$ and $N(r)$ are regular at the center.
The curvature scalars diverge at the center, the singularity
at the center is a mild one, it is a conical singularity.
The asymptotic mass $M$ of the object is presented by the integral
\begin{equation}
M =  \frac{|Q|}{4\sqrt{G}} \int^{\infty}_0 d \xi \left[
\frac{1}{\sigma(\xi)}
- \xi \frac{\sigma^{\prime}(\xi)}{\sigma^2(\xi)} - \frac{1}{2}
\sigma(\xi)\left(\xi^2 + 3 - \sqrt{\xi^4 + 2 \xi^2 +5}
\right) \right]
\,.
\label{mass2}
\end{equation}
Numerical calculations give the value
$M \simeq 0.442\,\frac{|Q|}{\sqrt{G}}$.
This is the illustration of the hypothesis that the non-minimal coupling of the electric field to the self-gravity field
can eliminate the singularity at the center of the object.

\subsubsection{Example of Regular Solution with pp-Wave Symmetry}

Now we consider a non-minimal pp-wave model, i.e., the model for which the metric functions $L$ and $\beta$, the axion field $\phi$, the electromagnetic field potentials ${\cal A}_2$ and ${\cal A}_2$ are the functions of the retarded time only (see \cite{BWTN2010} for details). Let us put $L(u) \equiv 1$.
In the standard theory of the gravity wave propagation there are no solutions with constant background factor $L(u)$;
moreover, the moment $u=u^{*}$ exists, for which $L(u^{*})=0$, i.e., the metric degenerates. The solution with $L \equiv 1$ can be indicated as the regular one, since
$det(g_{ik})= -L^4 \equiv -1$ and it can not vanish. For given solution for $L$, the equation for
$\beta$ reduces to
\begin{equation}
{-}\frac{2}{\kappa}  \left(\beta^{\prime}\right)^2 \left[1 {+}
\kappa \eta_{({\rm A})} \phi^2 \right] {=}
\left(\phi^{\prime}\right)^2 {+} \eta_{({\rm A})}
\left(\phi^2\right)^{\prime \prime} {+}
 \left({\cal A}^{\prime}_2 e^{{-}\beta}\right)^2 {+} \left({\cal
A}^{\prime}_3 e^{\beta}\right)^2   \,, \label{e358}
\end{equation}
where the prime denotes the derivative with respect to the retarded time $u$.
When $\eta_{({\rm A})}=0$ there are no real solutions of this
equation, but such a possibility appears in the non-minimal case.
We consider only one example of the exact regular models, it is
characterized by
\begin{equation}
 \phi = \phi_0  \,,  \quad {\cal A}_2(u)= {\cal
A}_2(0) e^{\beta(u)} \,, \quad {\cal A}_3(u)= {\cal A}_3(0) e^{-
\beta(u)} \,, \label{e0358}
\end{equation}
and is possible, when $\eta_{({\rm A})}<0 $ and
\begin{equation}
\kappa \phi^2_0 \ |\eta_{({\rm A})}| = 1 + \frac{\kappa}{2}\left[ {\cal
A}^2_2(0) + {\cal A}^2_3(0)\right] \,. \label{e243}
\end{equation}
For this specific solution the function $\beta(u)$ is arbitrary, and we suggested to use the
periodic finite function
\begin{equation}
\beta(u) = \frac{1}{2} \beta_{({\rm max })} (1-\cos{2\lambda u})
\,, \quad \beta(0) = 0 \,, \quad \beta^{\prime}(0) = 0 \,.
\label{e343}
\end{equation}
The metric for this non-minimal model is regular and periodic
\begin{equation}
\mbox{d}s^{2} {=} 2\mbox{d}u\mbox{d}v {-} \left\{\exp
\left[2\beta_{({\rm max })}\sin^2{\lambda u}\right]
(\mbox{d}x^2)^2 {+} \exp \left[{-}2\beta_{({\rm max
})}\sin^2{\lambda u}\right] (\mbox{d}x^3)^2 \right\} \,,
\label{30}
\end{equation}
the potentials of the electromagnetic field and their derivatives
are also periodic and regular. One can state that this regularity became possible due to non-minimal interaction of photons with the Dark Fluid via DM constituent.

\section{Model 10: Electromagnetic Interactions Induced by the Dark Fluid in a Plasma with Cooperative Field}

The Model 10 describes the example of indirect coupling of photons to the  Dark Fluid, mediated by  an ordinary matter, containing electric charges.
For the illustration we have chosen  the collisionless relativistic multi-component plasma, which is electro-neutral as a whole. We assume that the plasma particles interact
by cooperative electromagnetic field (the so-called Vlasov field), and this cooperative electromagnetic field is non-minimally coupled to the DM component of the Dark Fluid.
Mathematically, we use the axionic extension of the Einstein-Maxwell-Vlasov model (see \cite{BMZ2014A,EMV2,EMV3,EMV4} for details).

\subsection{Axionic Extension of the Einstein-Maxwell-Vlasov Model}

\subsubsection{The Extended Kinetic Equation}

The kinetic equation for a relativistic collisionless Vlasov's plasma can be written as follows:
\begin{equation}
\frac{p^i}{m_{({\rm a})}} \left[\frac{\partial}{\partial x^i} {-} \Gamma^k_{il}p^l \frac{\partial}{\partial p^k}\right]f_{({\rm a})} {+}
\frac{\partial}{\partial p^k} \left[ {\cal F}^k_{({\rm a})}f_{({\rm a})}\right] {=} 0
\,.
\label{k3}
\end{equation}
Here  the quantities $f_{({\rm a})}$ are the distribution functions of the particles of the sort ${({\rm
a})}$; $p^i$ is the momentum four-vector of the particle with the mass $m_{({\rm
a})}$. The term ${\cal F}^k_{({\rm a})}$ denotes a force acting on the charged particle in the axionically active plasma.
This force splits into the
standard Lorentz force linear in the particle momentum
four-vector, and the force ${\cal R}^i_{({\rm a})}$ induced by the
axion field
\begin{equation}
{\cal F}^i_{({\rm a})} = \frac{1}{m_{({\rm a})}}\left[ e_{({\rm a})} F^i_{\ s} p^s
+{\cal R}^i_{({\rm a})} \right] \,. \label{1check3}
\end{equation}
According to the kinetic theory the electric current four-vector $J^i$ contains the linear combination of first
moments of the distribution functions
\begin{equation}
J^i = \sum_{({\rm a})} e_{({\rm a})} \int dP  f_{({\rm a})} \ p^i \,.
\label{k2}
\end{equation}
The quantity $e_{({\rm a})}$ is the electric charge of the particle
of the sort ${({\rm a})}$; $dP= \sqrt{-g} \ d^4p$ is the invariant
integration volume in the momentum four-dimensional space. The
stress-energy tensor of the particles is presented by the second
moment of the distribution function
\begin{equation}
{\cal T}^{ik} = \sum_{({\rm a})} {\cal T}^{ik}_{({\rm a})} = \sum_{({\rm a})} \int dP  f_{({\rm a})} p^i p^k \,. \label{01EineqMIN}
\end{equation}
The particle momentum four-vector is normalized ($g_{ik} p^i p^k {=} m^2_{({\rm
a})}$), thus the trace of the stress-energy tensor
\begin{equation}
{\cal T} = \sum_{({\rm a})}{\cal T}_{({\rm a})} = g_{ik} {\cal T}^{ik}  = \sum_{({\rm a})} m^2_{({\rm
a})} \int dP  f_{({\rm a})}   \label{0145}
\end{equation}
is presented by the moment of zero order.

\subsubsection{Extended Non-minimal Equations of Axion Electrodynamics}

The cooperative electromagnetic field in plasma, which is described by the Maxwell tensor $F_{ik}$ entering the force ${\cal F}^k_{({\rm a})}$ (see (\ref{1check3})),
satisfies the electrodynamic equations in the integro-differential form
\begin{gather}
\nabla_k \left\{\eta_{(1)}  \nabla_m \phi R^{m[i} \nabla^{k]}\phi  + F^{ik} + {\cal R}^{ikmn} F_{mn} + \phi \left[ \Fst^{ik} + {\chi}^{ikmn}_{({\rm Axion})}
\Fst_{mn} \right] \right\} = {}\nonumber \\
{} = -  4\pi \sum_{({\rm a})} e_{({\rm a})} \int dP  f_{({\rm a})} \ p^i \,. \label{Vlasov2}
\end{gather}
As it was in the Model 9, the left-hand side of this equation includes terms describing the non-minimal mechanism of the Dark Fluid coupling to photons. The electric current in the right-hand side of the equation (\ref{Vlasov2}) describes the coupling mediated by the electrically neutral plasma.

\subsubsection{Extended Equation for the Pseudoscalar Field}

The non-minimally extended master equation for the pseudoscalar $\phi$
takes now the form
\begin{gather}
\nabla_m \left[ \left( \xi g^{mn} {+} \Re^{mn}_{({\rm A})} \right)
\nabla_n \phi \right] {+} \left[m^2_{({\rm A })} {+}
V^{\prime}(\phi^2){+} \eta_{({\rm A})} R \right] \phi = {}\nonumber \\
{}= - \frac{1}{\Psi^2_0}\left[\sum_{({\rm a})} \int dP f_{({\rm
a})} {\cal G}_{({\rm a})} {+} \frac{1}{4} \Fst_{mn}\left(
F^{mn}{+} {\chi}^{ikmn}_{({\rm A})} F_{ik} \right) \right]\,,
\label{eqaxi1}
\end{gather}
where $\xi=\pm 1$; $\Re^{mn}_{({\rm A})}$ and ${\chi}^{ikmn}_{({\rm A})}$ are
given by (\ref{sus3}) and (\ref{sus2}), respectively.
The pseudoscalar source
\begin{equation}
{\cal J} = \sum_{({\rm a})} \int dP  f_{({\rm a})} {\cal G}_{({\rm a})}   \label{j1}
\end{equation}
can be modeled by the pseudoscalar quantity ${\cal G}_{({\rm a})}$, which admits the decomposition
\begin{equation}
{\cal G}_{({\rm a})} = \alpha_{({\rm a})} \phi
+ \beta_{({\rm a})} p^k \nabla_k \phi
+ \gamma_{({\rm a})} p^k F^m_{\ k} \nabla_m \phi
+ ... \,. \label{j5}
\end{equation}
The phenomenological coefficients in this decomposition can be reconstructed using the compatibility conditions.

\subsubsection{Extended Gravity Field Equations and Reconstruction of the Effective Force Using the Compatibility Conditions}

The extension of the master equations for the gravitational field can be made by adding the source term (\ref{01EineqMIN}), where $f_{({\rm a})}$ are the solutions of the corresponding kinetic equations (\ref{k3}) with the forces containing $F^{ik}$, the solution to the electrodynamic equations (\ref{Vlasov2}).
According to the Bianchi identities, the total stress-energy tensor of the non-minimally interacting system is divergence-free. Since the divergence of the stress-energy tensor of plasma particles can be calculated as
\begin{gather}
\nabla_k {\cal T}^{ik} {=} \sum_{({\rm a})} m_{({\rm a})} \int dP f_{({\rm a})} {\cal
F}^i_{({\rm a})}\,, \label{check1}
\end{gather}
one can check directly (see \cite{BMZ2014A}) that the compatibility conditions are reduced to the relationships
\begin{gather}
\sum_{({\rm a})} \int dP f_{({\rm a})} \left[ {\cal R}^i_{({\rm
a})} - {\cal G}_{({\rm a})} \nabla^i \phi \right] = - \nabla_k T^{({\rm DE})ik}\,.
\label{check2}
\end{gather}
In other words, the additional force ${\cal R}^i_{({\rm
a})}$, which acts on the plasma particles due to the Dark Fluid influence, can be divided into two parts. First, the contribution of the axionic Dark Matter is predetermined by the structure of the quantity ${\cal G}_{({\rm a})}$ decomposed as (\ref{j5}). Second, the contribution of the Dark Energy is connected with the structure of the term $\nabla_k T^{({\rm DE})ik}$.

For an illustration, we consider below the case, when the force is orthogonal to the
particle four-momentum, thus providing the particle mass conservation:
\begin{equation}
{\cal R}^i_{({\rm a})} = \left[\delta^i_k \ (p_s p^s) - p^i p_k
\right] \nu_{({\rm a})}  {\cal B}^k  \,, \quad {\cal R}^i_{({\rm
a})} p_i \equiv 0 \,,\label{check4}
\end{equation}
where $\nu_{({\rm a})}$ are phenomenological constants. The unknown four-vector  ${\cal B}^k $ can be reconstructed as
\begin{equation}
{\cal B}^k = \tilde{{\cal S}}^k_l  \ \left[{\cal J} \nabla^l \phi - \nabla_m T^{({\rm DE})lm} \right] = \tilde{{\cal S}}^k_l  \ \left[{\cal J} \nabla^l \phi - \left(D+\Theta \right)(U^l W_{({\rm DE})}) - \nabla_m {\cal P}^{lm} \right] \,, \label{check7}
\end{equation}
where ${\cal J}$ is represented by (\ref{j1}) with  (\ref{j5}), and $\tilde{{\cal S}}^k_l$
is reciprocal to ${\cal S}^i_k $:
\begin{equation}
\tilde{{\cal S}}^i_k {\cal S}^k_l =
\delta^i_l \,, \quad {\cal S}^i_k \equiv \left[\delta^i_k \ S_m^m - S^i_k \right] \,, \quad S^i_k  = \sum_{({\rm a})} \nu_{({\rm a})} {\cal T}^i_{k ({\rm a})}\,. \label{check55}
\end{equation}
Thus, the compatibility conditions are satisfied, the model as a whole is self-consistent, and the ponderomotive force, which acts on the plasma particles from the Dark Fluid, is describes by
(\ref{check4}) with  (\ref{check7}), (\ref{check55}).

\subsection{First Application:  Propagation of Electromagnetic Waves in an Axionically Active Ultrarelativistic Plasma Non-minimally Coupled to Gravity in a de Sitter Background}

\subsubsection{Dispersion Relations}

Based on approach standard for the plasma theory, we consider the state of plasma perturbed by a local
variation of electric charge, assuming that the
distribution function obtains a small variation $f_{({\rm a})} \to f^{(0)}_{({\rm a})} {+}  \delta f_{({\rm a})}$,
and, assuming that the cooperative electromagnetic field was vanishing in the unperturbed state (in the isotropic homogeneous de Sitter cosmological background).
Then, as it was shown in \cite{EMV3}, the dispersion relation for longitudinal electric waves in axionically active plasma non-minimally coupled to gravity is extended as follows:
\begin{equation}
\varepsilon_{||}(\Omega, k_{\alpha}) = - 2K_1 \,,
\label{longDR}
\end{equation}
where the new non-minimal parameter
\begin{equation}
K_1 \equiv -H^2(6q_1{+}3q_2{+}q_3) \label{eld777}
\end{equation}
appears in the right-hand side of this equation, with the  Hubble constant $H$ and non-minimal coupling constants $q_1$, $q_2$, $q_3$ (see (\ref{sus1}) for the susceptibility tensor
${\cal R}^{ikmn}$).
The longitudinal dielectric permittivity
\begin{equation}
\varepsilon_{||} \equiv 1 + \frac{4\pi}{k^2}\sum_{({\rm a})}
e_{({\rm a})}^2 \int \frac{d_3q\, k_\alpha q^\alpha}{(q\Omega-k_\beta q^\beta)}\cdot\frac{df^{(0)}_{({\rm a})}}{dq}
\label{long4}
\end{equation}
is the function of complex frequency $\Omega = \omega + i \gamma$, and of components of a real wave three-vector $k_{\alpha}$. Since the illustration is prepared for the ultrarelativistic plasma,
we assume that the particle energy is given by $q$, defined by $q^2 = - q_{\alpha} q^{\alpha}$ via the particle three momentum $q^{\alpha}$.

The dispersion relation for transversal electromagnetic waves in axionically active plasma non-minimally coupled to gravity
can be written in the form
\begin{gather}
\left[\varepsilon_\bot{+}2K_1{-}\frac{(1{+}2K_1)k^2}{\Omega^2}\right]
\left[\left(\varepsilon_\bot {+}2K_1 {-}
\frac{(1{+}2K_1)k^2}{\Omega^2}\right)^2 -\frac{\nu^2(1{+}2K_2)^2k^2}{\Omega^4}\right]= 0 \,,
\label{trans1}
\end{gather}
where a new non-minimal constant
\begin{equation}
K_2 \equiv -2H^2(3Q_1{-}Q_3)  \label{eld7}
\end{equation}
contains the non-minimal coupling constants $Q_1$, $Q_2$ entering the susceptibility tensor $\chi^{ikmn}_{({\rm Axion})}$ (\ref{sus2}).
The transversal permittivity scalar $\varepsilon_{\bot}$ is given by the integral
\begin{equation}
\varepsilon_\bot \equiv  1+ \frac{2\pi }{k}\sum_{({\rm a})}
e_{({\rm a})}^2 \int \frac{q d_3q\,}{(q\Omega-k_\beta q^\beta)}\left[\frac{k}{\Omega}- \frac{k_\beta q^\beta}{kq} \right]\cdot\frac{df^{(0)}_{({\rm a})}}{dq}
\label{e1}
\end{equation}
and the parameter $\nu$ is proportional to the time derivative of the axion field.
The dispersion relation for longitudinal electric waves does not contain information about axion field; only  transversal electromagnetic waves are influenced by the axionic Dark Matter.

\subsubsection{Non-minimal Coupling of Transversal Plasma Waves to a Stationary Axionic Dark Matter}

When the Dark Matter is stationary, $ \dot{\phi}=0$, thus $\nu=0$ and dispersion relations (\ref{trans1}) reduce to
\begin{gather}
\varepsilon_{\bot}-1 = (1{+}2K_1)\left(\frac{k^2}{\Omega^2}-1 \right)\,.
\label{trans12}
\end{gather}
When  $1{+}2K_1\geq 0$  (clearly, the classical case $K_1=0$ is also included), there are no solutions of (\ref{trans12}) with $\omega<k$ (see \cite{EMV2,EMV3} for details). This means that in such stationary Dark Matter the transversal electromagnetic waves propagate with phase velocity exceeding the speed of light in vacuum, $\frac{\omega}{k}>1$; these are running waves without damping.
When $1{+}2K_1 < 0$, or in other words $6q_1{+}3q_2{+}q_3>\frac{1}{2H^2}$, there exist the solution with $\frac{\omega}{k}<1$; the transversal electromagnetic waves move with phase velocity less than the speed of light in vacuum, thus the resonant interaction with co-moving charged particles is possible, and the Landau damping leads to the wave attenuation (see \cite{EMV2,EMV3}).

\subsubsection{Non-minimal Coupling of Transversal Plasma Waves to a Non-stationary Axionic Dark Matter}

Let us consider the non-minimal cosmological model, for which the trace of the susceptibility tensor ${\cal R}^{ikmn}$ vanishes, i.e., $6q_1{+}3q_2{+}q_3=0$, and thus $K_1=0$.
We introduce the notation $p \equiv \nu (1{+}2K_2)$, and consider the dispersion equation for the transversal electromagnetic waves in plasma
\begin{equation}
\varepsilon_\bot = \frac{k^2\pm pk}{\Omega^2} \,.
\label{transDR}
\end{equation}
Plus and minus in this formula relate to the transversal electromagnetic waves with left-hand and right-hand polarization rotation, respectively. In this sense, when $p \neq 0$, we deal with an
axionically active plasma, which produces the effect of optical activity in plasma, similar to the optical activity effect in axionic vacuum \cite{Itin2}. As it was shown in \cite{EMV3}, the phase velocity of transversal waves can be less than  speed of light in vacuum,
$\frac{\omega}{k}<1$, when
\begin{gather}
1> \frac{p}{k}> \frac{3}{2k^2 r^2_{({\rm D})}}\,,
\label{trans123}
\end{gather}
where the Debye radius $r_{({\rm D})}$ is defined standardly as
\begin{gather}
\frac{1}{r^2_{({\rm D})}} = \frac{4\pi}{3} \sum_{({\rm a})} \frac{e^2_{({\rm a})} N_{({\rm a})}}{k_{\rm B} T_{({\rm a})}}
\label{trans125}
\end{gather}
with the charge $e_{({\rm a})}$, number density $N_{({\rm a})}$ and temperature $T_{({\rm a})}$, indicated by the sort index $({\rm a})$ ($k_{\rm B}$ is the Boltzmann constant).

\subsection{Second Application:  Cosmological Electric Field Induced by Axionic Dark Matter in a Bianchi-I Model with Magnetic Field}

The second application of the Einstein-Maxwell-Vlasov-axion model relates to the Bianchi-I anisotropic homogeneous cosmological model with local rotational isotropy; for this purpose we use  the metric
\begin{equation}
ds^2 = dt^2-a^2(t)({dx^1}^2+{dx^2}^2)-c^2(t){dx^3}^2 \,,
\label{1gr}
\end{equation}
and consider the scale factors $a(t)$ and $c(t)$ to be functions of cosmological time $t$. We assume that the axion field, as well as the parallel electric and magnetic fields inherit the
space-time symmetry: these quantities depend on time only, $B^3(t)\neq 0$ and $E^3(t)\neq 0$.

Especial interest to this model is motivated by the following reasoning. The magnetic field $B^i$ in the non-stationary axionic environment ($\dot{\phi}\neq 0$) is known to produce electric field $E^i$ parallel to the magnetic one, and the proportionality coefficient is linear in the axion field (see, e.g., \cite{BBT2012,EMV4}).
The corresponding pseudoscalar $E^i B_i$, which forms the electromagnetic source in the right-hand side of the master equation for the axion field (\ref{eqaxi1}), happens to be linear in $\phi$ and quadratic in the initial magnetic field. This means that the backreaction of the electromagnetic field on the axion field can be described in terms of effective pseudoscalar mass. Let us illustrate this sentence by the master equations obtained in \cite{BBT2012} for the Bianchi-I model:
\begin{equation}
F_{12}(t) = F_{12}(0) \,, \quad F^{30}(t) =  \frac{F_{12}}{a^2(t)c(t)} \phi(t) \,.
\label{cosm5}
\end{equation}
The corresponding physical components of the magnetic $B(t)$ and electric $E(t)$ fields are given by
\begin{equation}
B(t) \equiv \sqrt{F_{12}F^{12}}= \frac{F_{12}}{a^2(t)} \,, \quad E(t) \equiv \sqrt{-F_{30}F^{30}}= \frac{F_{12} \phi(t)}{a^2(t)} = B(t)\phi(t) \,,
\label{cosm555}
\end{equation}
and the master equation for the axion field takes the form
\begin{equation}
\ddot \phi+ \left(\frac{2\dot a}{a}+\frac{\dot
c}{c}\right) \dot \phi+ \phi \left[m^2_{({\rm A })} \pm
\frac{F^2_{12}}{\Psi^2_0 a^4} \right] = 0 \,.
\label{cosm9}
\end{equation}
The sign plus in the equation (\ref{cosm9}) relates to a canonic axionic Dark Matter, while the sign minus appears, when the Dark Matter is phantom-like, or in other words,
it can be described by the Lagrangian with negative kinetic part. In the first case, the square of effective pseudoscalar mass
\begin{equation}
M^2_{({\rm A })} = m^2_{({\rm A })} +
\frac{F^2_{12}}{\Psi^2_0 a^4}
\label{cosm98}
\end{equation}
is positively defined, and evolution of the axion field is characterized by a quasi-oscillatory regime.
For the second case there exists a situation, when  the quantity
\begin{equation}
m^2_{({\rm A })} -
\frac{F^2_{12}}{\Psi^2_0 a^4}  = - \zeta^2
\label{cosm97}
\end{equation}
is negative, and thus the axionic field grows in an inflationary-type manner (see \cite{BBT2012}).

The growing electric field inevitably polarizes the multi-component plasma, and one can expect that cooperative Vlasov's electric field in plasma will counteract the external axionically induced electric field. This counteraction produces a new oscillatory regime with another set of eigen-frequencies.
Indeed, the Maxwell equation for the electric field $E(t)$ can be reduced to the equation for the potential $A_3(t)$:
\begin{equation}
{\ddot{A}}_3 + {\dot{A}}_3 \left(\frac{2\dot
a}{a}-\frac{\dot c}{c}\right) + \Omega^2_{\rm L} A_3 = F_{12} \dot{\phi} \left(\frac{c}{a^2} \right)  \,,
\label{Maxw44}
\end{equation}
where $F^{30}=\frac{1}{c^2(t)} {\dot{A}}_3$. The quantity
$\Omega_{\rm L}$ is the Langmuir frequency in the relativistic plasma; its non-relativistic analog is
\begin{equation}
\Omega^2_{\rm L} = 4\pi \sum_{({\rm a})} \frac{e^2_{({\rm a})} N_{({\rm a})}}{m_{({\rm a})}} \,.
\label{Maxw445}
\end{equation}
Combined with the modified equation for the axion field
\begin{gather}
\label{bas10b}
\ddot \phi+ \left(\frac{2\dot
a}{a}+\frac{\dot c}{c}\right) \dot \phi + m^2_{({\rm A })}\phi = {\dot{A}}_3 \ \frac{F_{12}}{\Psi^2_0 a^2 c}\,,
\end{gather}
the equation (\ref{Maxw44}) gives the set of equations for electro-axionic oscillations. The spectrum of these oscillations is studied in details in \cite{EMV4}. Here we would like to display only two important details. First, the equation describing the frequency $\Omega$ of the  electro-axionic oscillations is given by the following biquadratic equation:
\begin{gather}
\Omega^4 - \Omega^2 \left(\Omega^2_{\rm L} + m^2_{({\rm
A})}-\Omega^2_{\rm B}\right)+m^2_{({\rm A })}\Omega^2_{\rm L} =0\,,
\label{f6}
\end{gather}
where the auxiliary constant $\Omega^2_{\rm B}= \frac{F_{12}^2}{a^4(t_0)\Psi^2_0}$ is introduced.
Second, using the so-called combination frequencies:
\begin{gather}\label{odr1}
\Omega_{\pm} \equiv \Omega_{\rm L} \pm m_{({\rm A })} \,,
\end{gather}
we can write the real solutions of this biquadratic equation as
\begin{gather}\label{odr966}
\Omega = \pm \frac12 \left[\sqrt{\Omega^2_{+} - \Omega^2_{\rm B}} \pm  \sqrt{\Omega^2_{-} - \Omega^2_{\rm B}} \right]\,.
\end{gather}
The presented roots are real, when $\Omega^2_{\rm B}<\Omega^2_{-}$.

To sum up the results, one can say that the discussed model describes the interaction of the following quartet: first, the anisotropic homogeneous cosmological gravitational field; second the axionic Dark Matter; third, the global Longitudinal Magneto-Electric Cluster; fourth, the cooperative Vlasov's electric field in plasma. It is a sophisticated four-level interaction, and it is interesting that an oscillatory regime in this system is possible, the frequencies of which are presented by (\ref{odr966}).

\section{Conclusions}

The main idea of this work is to show that the cosmic Dark Fluid can be considered as an electromagnetically active medium, which indirectly affects on electromagnetic fields of all types.
There are many excellent reviews (see Introduction and references therein), which are focused on detailed description of cosmological aspects of evolution of the Dark Energy and Dark Matter coupled by the gravitational field. Also, one can find many reviews describing properties of uncharged particles, which (hypothetically) form the Dark Matter (axions, axion-like particles, WIMPs, etc.). The presented review is focused on electrodynamic aspects of the Dark Fluid evolution, and thus supplements mentioned surveys. In other words, our goal is to look on the problem of the Dark Matter and Dark Energy identification from the point of view of response of an electromagnetic field on the Dark Fluid influence.

Based on description of ten models of coupling of electromagnetic fields and Dark Fluid, we can formulate three typical consequences of such interactions.

\vspace{3mm}
\noindent
1. The axionic Dark Matter, the first (pseudoscalar) constituent of the Dark Fluid, provides the global physical system to become the chiral one. One can distinguish three symptoms of chirality provided by the Dark Fluid.

\noindent
{\em (i)} The first symptom of chirality is the effect of optical activity of the axionically active vacuum, plasma, dielectric media. This effect reveals itself in a polarization rotation of running and standing electromagnetic waves.

\noindent
{\em (ii)} The second symptom of chirality is the generation of specific Longitudinal Magneto-Electric Clusters. The term "Longitudinal" means that  due to the coupling to the axionic Dark Matter, the magnetic field generates an electric field parallel to the initial magnetic field. This effect is typical for the axion electrodynamics, but does not appear in the standard Faraday-Maxwell electrodynamics. The Longitudinal Magneto-Electric Clusters are shown to appear in models for anisotropic cosmology, in static models with spherical symmetry, in models with pp-wave symmetry. Such axionically produced Longitudinal Clusters can appear ((e.g., in the Earth Ionosphere)) as specific oscillations, in which time-depending magnetic and electric fields are collinear.

\noindent
{\em (iii)} The third symptom of chirality is connected with a specific contribution into the dynamo-optical phenomena, activated by the axionic Dark Matter. Such effects can appear, when the Dark Fluid moves non-uniformly, and the pseudoscalar (axion) field is non-stationary or inhomogeneous.

\vspace{3mm}
\noindent
2. The second typical consequence of photon coupling to the Dark Fluid is the generation of anomalous response of electrodynamic systems, in appropriate physical conditions. We have shown that the interaction of axionic Dark Matter with initially constant magnetic field in the field of gravitational waves, produces anomalously amplified electric field. Anomalies can appear in anisotropic expanding Universe with magnetic field, as well as, a static anomaly can be formed in the vicinity of axionic monopoles and stars.

\vspace{3mm}
\noindent
3. In the cosmological context, the interaction of photons with Dark Fluid can organize specific Dark Epochs in the Universe history, during which the effective refraction index of the cosmic medium becomes an imaginary quantity. This means that electromagnetic waves can not propagate during such Dark Epochs, and the corresponding electromagnetic energy-information transfer is stopped.  We have shown that the formation of Dark Epochs can be caused by both constituents of the Dark Fluid: by the axionic Dark Matter (e.g., in the model of gradient-type extension of axion electrodynamics), and by the non-stationary Dark Energy (e.g., in the model of striction-type activity, and in the model of Archimedean-type coupling to the Dark Matter).

To conclude, we would like to emphasize that the results described in the review have a status of theoretical findings. The question arises:
is there experimental information available that can allow us to select one, two or three of these ten models, and to indicate them as preferable for the Dark Fluid Electrodynamics?
We think that a part of the necessary information already exists, but is still in a hidden form. For instance, if the Universe indeed has passed through dark epochs,
the CMB data accumulated in the WMAP archive (see, e.g., \cite{WMAP}) can contain fingerprints of such events; we believe that a special procedure of data processing could reveal
these fingerprints of the Dark Energy influence, thus providing constraints for the coupling constants listed in the review. Long-term records of variations of the electric and magnetic fields in the Earth Ionosphere give another source of hidden information about the coupling of photons to the axionic Dark Matter. In \cite{Vladimir} one can find first results of the corresponding data processing aimed to verify the predictions about  axionically induced magneto-electric oscillations in the Earth Ionosphere (see \cite{BG2013} for details).
New information about phenomenological parameters, appeared in the Dark Fluid electrodynamics, can be obtained in experiments concerning the axionically induced spin precession (see, e.g., \cite{SA1,SA2,BP2015}.

\acknowledgments{The work was supported by Russian Science Foundation (Project No. 16-12-10401).}

\end{document}